\documentclass[12pt,preprint]{aastex}
\usepackage{enumerate}

\def\be{\begin{equation}}
\def\ee{\end{equation}}
\def\cms{{\rm cm\,s^{-1}}}
\def\kms{{\rm km\,s^{-1}}}
\def\ms{{\rm m\,s^{-1}}}

\def\AU{{\rm AU}}
\def\msun{{M_\odot}}

\begin{document}
\title{The Rossiter-McLaughlin effect for exomoons or binary planets}
\author{Quntao Zhuang, Xun Gao, \& Qingjuan Yu\thanks{Email: yuqj@pku.edu.cn}
}
\affil{Kavli Institute for Astronomy and Astrophysics, and School of Physics,
Peking University, Beijing, 100871, China }
\begin{abstract}
In this paper we study possible signatures of binary planets or exomoons on
the Rossiter-McLaughlin (R-M) effect. Our analyses show that the R-M effect for
a binary planet or exomoon during its complete transit phase can be divided into
two parts. The first is the conventional one similar to the R-M effect from the
transit of a single planet, of which the mass and the projected area are the
combinations of the binary components; and the second is caused by the orbital rotation of the
binary components, which may add a sine- or linear-mode deviation to the
stellar radial velocity curve. We find that the latter effect can be up to
several or several ten $\ms$.  By doing numerical simulations as well as
analytical analyses, we illustrate that the distribution and dispersion of the
latter effects obtained from multiple transit events can be used to constrain
the dynamical configuration of the binary planet, such as, how the inner orbit
of the binary planet is inclined to its orbit rotating around the central star.
We find that the signatures caused by the orbital rotation of the binary
components are more likely to be revealed if the two components of binary
planet have different masses and mass densities, especially if the heavy one
has a high mass density and the light one has a low density.  Similar signature
on the R-M effect may also be revealed in a hierarchical triple star system
containing a dark compact binary and a tertiary star.
\end{abstract}

\section{Introduction}\label{sec:intro}

Planets are not alone. For example, most of the planets in the Solar System
have surrounding satellites or moons; not only some asteroids, but also many of
the Kuiper Belt objects (e.g., Pluto and Charon) are in binaries; and recent
observations have also revealed miscellaneous exoplanetary worlds, including
multi-planetary systems (e.g., Kepler-11, HD 10180, Gliese 581;
\citealt{Lissauer11,Lovis10,V10,F11}). Although the existence of exomoons or binary
exoplanets have not been detected, various detection methods have been
proposed, such as through transit light curves, transit timing and duration
variations, direct imaging, microlensing, or Doppler spectroscopy of the host
planet (e.g., \citealt{Kipping09,Simon10,Simon12,SS99}; and references therein). In
this paper, we investigate possible observational signatures that a binary
exoplanet system (if any) or an exoplanet plus one moon system would have on
the Rossiter-McLaughlin (R-M) effect \citep{M24,R24} and how the orbital
configurations of the system could be revealed through the signatures. For
simplicity, below we also call the one exoplanet plus one moon system as a
binary planet, although the two components have very different masses.
Existence of binary planets and statistics on their dynamical configurations
should shed new light on the formation and evolution of planetary systems and
the search for a habitable world.

When a planet transits in front of a rotating star, it blocks part of the light
emitted from the stellar surface, and the blocked region shifts with the
transiting of the planet. As different parts of the stellar surface may have
different line-of-sight velocities to the observer, the shifting of the blocked
region results in either blueshift or redshift of observed stellar spectral
lines and further the deviation of the inferred radial velocity of the stellar
motion (i.e., the R-M effect).\footnote{Note that 
extraction of the velocity deviation from stellar spectrum line
profiles involves some detailed techniques in the modeling of the effect
(e.g., \citealt{H11,A07}.)}
The deviation of the stellar radial velocity
provides a way to measure the misalignment between the stellar spin and the
planetary orbital angular momentum, and recent measurements of the R-M effects
have found that some exoplanets are on highly inclined orbits relative to the
spin of the star (e.g., retrograde or polar orbits; \citealt{Winn09, CC10}). For the
similar reasons, if the transiting planet is a binary, the binary with
different physical properties and orbital configurations (e.g., radii,
inclination) may block the stellar surface in different ways, and hence the
shift of stellar spectral lines and the deviation of its radial velocity curve
may have different signatures from those expected from the transit of one
single planet.  The gravitation from a binary planet is also different from
that from one single planet with the same total mass, but the resulted
deviation in the dynamical motion of the star is generally small within each
transit duration (see justification in Section~\ref{subsec:observation} below).
In this paper, we isolate and illustrate the effects on the observed stellar
radial velocity due to the different light blocking ways.  

Most of the modeling of the R-M effect for planetary systems were developed for
a single planet rotating around a star. The R-M effect for exomoon systems was
numerically modeled by \citet{Simon10}, where the exomoon has a much smaller
mass and radius than its host planet and the moon effects are modeled as a
small perturbation added to the effect caused by a single planet. Their
modeling includes the fitting to the numerous dynamical parameters of the
systems, including instantaneously and fast evolving ones. In this paper we
present a comprehensive analysis and investigation on the R-M effect for a
binary planet system, where the satellite mass is not limited to be small but
can be comparable to the host planet mass.  We include the effects from the
dynamical evolution of the binary planet system during multiple transits and
their different dynamical configurations. Our detailed treatments average out
the effects from some instantaneously changing dynamical angles and include the
evolutionary patterns of those relatively fast changing ones, so that we can
focus on the effects from different inner orbital inclinations of the binary
planet relative to its orbit rotating around the central star.

The paper is organized as follows. In Section~\ref{sec:geometry}, we describe
the geometric configuration of the system to be studied in this paper (i.e., a
binary planet transiting in front of a star) and related dynamical
approximations.  In Section~\ref{sec:RM}, we investigate how the R-M effect is
affected by a binary planet and how the different orbital configurations could
be inferred from the deviation of the stellar radial velocity curves, together
with the transit of the stellar light curves. We explore the parameter space of
binary planets that are likely to be revealed in observations.  We also extend
the results to hierarchical triple star systems in which a binary star (e.g., a
compact stellar remnant plus a brown dwarf or planet) is transiting in front of
a tertiary star gravitationally bound to the system.
Section~\ref{sec:discussion} contains a summary and a discussion.

\section{Geometric configuration and dynamical description of the system}
\label{sec:geometry}

\begin{figure}
\epsscale{0.4}
\plotone{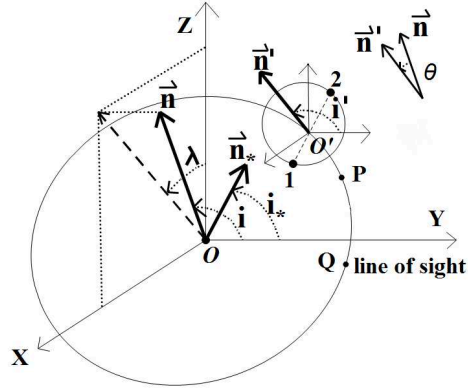}
\caption{Schematic diagram for the orbital configuration of the system in which
a binary planet is rotating around a star. The star is located at the origin
$O$. The two components of the binary planet are denoted by ``1'' and ``2'',
and their center of mass is labeled by $O'$. For brevity, the star and the
binary planet are all indicated by points, without illustrating their physical
radii in the figure. The observer is along the direction of the $y$-axis. 
The spin direction of the star $\vec{n}_*$ is located at the $y$-$z$ plane. 
The round curve centering at $O$ represents the orbit of the center of mass of
the binary planet rotating around the star. The $\vec{n}$ represents the
unit vector of the orbital angular momentum of the outer binary,
and the $\vec{n}'$ represents the unit vector of the orbital angular momentum
of the inner binary (binary planet). The angle between $\vec{n}$ and $\vec{n}'$
is defined by $\theta$ in Table~\ref{tab:para}.
Point $P$ is the pericenter of the orbital plane of the outer binary if the orbit is
eccentric, and point $Q$ is located on the intersection line of that orbital plane and
the plane where both the vector $\vec{n}$ and the $y$-axis are located. The angle
between $\protect\overrightarrow{OP}$ and $\protect\overrightarrow{OQ}$ ($\angle POQ$
not labeled in the figure for visual clarity) is used in equation (\ref{eq:g}) later.
For other labeled angles, see also Table~\ref{tab:para} and Section~\ref{sec:geometry}.
}
\label{fig:geom} \end{figure}

Consider that a binary planet is rotating around a star (see
Fig.~\ref{fig:geom}). For convenience, we shall call the binary planet as the
``inner'' binary, and call the star and the center of the mass of the binary
planet as the ``outer'' binary. We denote the mass and the radius of the star
as $m_*$ and $R_*$. The projected area of the star onto the sky is $A_*=\pi
R_*^2$, and it has a surface brightness of $I_*$. The angular velocity of the
stellar spin is $\Omega_*$. We denote the component masses of the inner binary
as $m_1$ and $m_2$ ($m_1\ge m_2$), and the component radii as $R_1$ and $R_2$,
respectively.  Here we have $m_1+m_2\ll m_*$. The projected areas of the two
components onto the sky are $A_1=\pi R_1^2$ and $A_2=\pi R_2^2$, respectively
(see the parameter list in Table~\ref{tab:para}). 

\begin{deluxetable}{lcc}
\tablecaption{List of parameters.}
\tablehead{\colhead{Object} & \colhead{Property} & \colhead{Symbol} }
\startdata
Central star &mass               & $m_*$\\
             &radius             & $R_*$ \\
             &sky-projected area     & $A_*$\\
             &surface brightness & $I_*$\\
             &unit vector of spin & $\vec{n}_*$\\
             &inclination of spin& $i_*$\\
             &spin angular velocity&$\Omega_*$\\
\hline
Binary planet &mass&$m_1, m_2$ \\
&radius&$R_1, R_2$\\
&sky-projected area&$A_1, A_2$\\
&mass density&$\rho_1, \rho_2$\\
\hline
Orbit of outer binary&semimajor axis&$a$\\
&eccentricity&$e$\\
&unit vector of angular momentum & $\vec{n}$\\
&angular velocity&$\omega$\\
&sky-projected angle between $\vec{n}$ and $\vec{n}_*$ & $\lambda$\\
&inclination to the line of sight&  $i$\\
\hline
Orbit of inner binary  &semimajor axis& $d$\\
&distance to center of mass &$d_1,d_2$\\
&eccentricity &$e'$\\
&unit vector of angular momentum & $\vec{n}'$\\
&angular velocity&$\omega'$\\
&sky-projected angle between $\vec{n}'$ and $\vec{n}_*$& $\lambda'$\\
&inclination to the line of sight & $i'$\\
\hline
Precession of $\vec{n}'$ around $\vec{n}$ & angle between $\vec{n}$ and $\vec{n}'$ & $\theta$\\
& angular velocity&$\Omega$\\
\enddata
\label{table:t1}
\tablecomments{See details in Section~\ref{sec:geometry} and Fig.~\ref{fig:geom}.}
\label{tab:para}
\end{deluxetable}

We describe the dynamical motion of the system through the two components: (i)
the orbital motion of the outer binary, and (ii) the orbital motion of the
inner binary. The orbital motion of the inner binary includes the precession of
its orbital angular momentum around the orbital angular momentum of the outer
binary, as described below.

The orbital configuration of the system is indicated in a reference
frame as shown in Figure~\ref{fig:geom}. 
In Figure~\ref{fig:geom}, the center of the star is located at the origin $O$.
The $y$-axis is directed toward the observer, and the $z$-axis is chosen so
that the stellar spin axis lies on the $y$-$z$ plane.  The inclination angle of
the stellar spin relative to the $y$-axis is denoted by $i_*$ ($0\le
i_*\le\pi$). The unit vector of the orbital angular momentum of the outer
binary is denoted by $\vec{n}$, and we define $\lambda$ by the angle between
the $z$-axis and the projected vector of $\vec{n}$ onto the $x$-$z$ plane
($0\le\lambda<2\pi$). The orbital inclination angle to the observer $i$ is
defined by the angle between $\vec{n}$ and the $y$-axis. Thus, we have
$\vec{n}=(\sin\lambda\sin i, \cos i, \cos\lambda\sin i)$. The outer binary has
a semimajor axis of $a$ and an angular velocity $\omega=[G(m_*+m_1+m_2)/a^3]^{1/2}$.
For simplicity, the eccentricity of the outer binary $e$ is assumed to be zero,
unless otherwise specified. 

Similarly, we denote the unit vector of the orbital angular momentum of the
inner binary by $\vec{n}'$, and we define $\lambda'$ by the angle between the
$z$-axis and the projected vector of $\vec{n}'$ onto the $x$-$z$ plane, and
define the orbital inclination angle $i'$ to the observer by the angle between
$\vec{n}'$ and the $y$-axis. We have $\vec{n}'=(\sin\lambda'\sin i', \cos i',
\cos\lambda'\sin i')$. The angle between $\vec{n}$ and $\vec{n}'$ is denoted by
\be
\theta\equiv\arccos(\vec{n}\cdot\vec{n}'), \quad 0\le\theta\le\pi.
\label{eq:theta}
\ee
The semimajor axis and angular velocity of the inner
binary are denoted by $d$ and $\omega'=[G(m_1+m_2)/d^3]^{1/2}$, respectively.
The eccentricity of the inner binary $e'$ is assumed to be zero, unless specially
discussed in some cases below.
Note that the semimajor axis of the inner binary is limited by the Hill radius $d_H$
and the Roche limit $d_R$, i.e.,
\begin{eqnarray}
d\la d_H\equiv a\left(\frac{m_1+m_2}{3m_*}\right)^{1/3}
\simeq 0.07\AU\left(\frac{a}{1\AU}\right)\left(\frac{m_1+m_2}{M_J}\right)^{1/3}\left(\frac{\msun}{m_*}\right)^{1/3}
\label{eq:Hill}
\end{eqnarray}
\be
d\ga d_R\equiv R_1\left(2\frac{\rho_1}{\rho_2}\right)^{1/3},
\label{eq:Roche}
\ee
where $M_\odot$ is the solar mass, $M_J$ is the Jupiter mass, $\rho_i$ ($=\frac{3m_i}{4\pi R_i^3}$, $i=1,2$)
is the mass density of each component of the inner binary.
An inner binary with a larger semimajor axis will be tidally broken up by the
gravitation from the star.

In general, the orbital parameters of the outer binary and the inner binary may evolve with
time under the three-body interactions of the star and the binary planet. But under
some conditions, their dynamical motion can be much simplified as follows.
\begin{itemize}
\item The orbital parameters of the outer binary ($a$, $e$, $\vec{n}$, $i$, $\lambda$) can
be approximately constant, if the semimajor axis of the inner binary is much smaller
than the Hill radius $d_H$.
\item The semimajor-axis of the inner binary $d$ can also be approximately constant if
it is much smaller than the Hill radius.
Both the inclination $i'$ and the projected angle $\lambda'$
of the inner binary may change with time due to the evolution of $\vec{n}'$.
The evolution of $\vec{n}'$ depends on the angle $\theta$ between the
orbital angular momenta of the inner and the outer binaries
(i.e., $\vec{n}'$ and $\vec{n}$; e.g., \citealt{Ford00}).
\begin{enumerate}[(a)]
\item If the angle $\theta$ or $180\degr-\theta$ is small (e.g., $\la 40\degr$,
the Kozai angle; \citealt{K62}), the $\vec{n}'$ precesses around $\vec{n}$
approximately at a constant angular velocity $\Omega\sim
\omega^2\cos\theta/\omega'$, and the angle $\theta$ can also be approximately
constant. The angles $i'$ and $\lambda'$ change periodically with the
precession of $\vec{n}'$.  As one transit duration $\delta t$ ($\sim
\omega^{-1}R_*/a$) is generally much shorter than the precession timescale of
$\vec{n}'$, the $\vec{n}'$ (and $i'$, $\lambda'$) can be approximated as
constant during each transit duration. 

\item If the angle $\theta$ is about in the range from $40\degr$ to $140\degr$,
the Kozai mechanism affects the evolution of the orbital parameters of the
inner binary, in which the angle $\theta$ and the eccentricity $e'$
exchange at a period $\sim 2\pi\omega'/\omega^2$ due to
angular momentum transfer and the conservation of the quantity
$C_K\equiv \sqrt{1-e'^2}\cos\theta$ (see an example shown in Fig.~\ref{fig:ftheta} below).  
During the oscillation, the eccentricity of the
inner binary can be induced to high values close to 1, and thus the inner binary is likely
to be destroyed by collision of its two components. If there exist other
moons in the system, one component of the binary is also likely to be ejected
from the system or collide with one of the moons.
The Jovian system is such an example influenced by the Kozai mechanism, and
almost all the inclinations of their moons are out of the angle range. Here we
ignore other effects from the planet (e.g., tides, the general relativistic
effect), which would limit the influence of the Kozai mechanism.  The
limitation would become relatively significant for a system with small $d$ but
large $a$. For example, Uranus is {\it farther} from the sun and its {\it inner}
moons (with $R_*/a\simeq 2.4\times 10^{-4}$ and
$a/d\sim 5\times10^3-6\times10^4$)
are on polar orbits relative to its orbital plane surrounding the sun
(e.g., \citealt{MD99}).

\end{enumerate}
Below we do not focus on case (b).
\end{itemize}

\section{Transit of a binary planet and its signatures on the R-M effect} \label{sec:RM}

\subsection{Transit light curves} \label{subsec:transit}

\begin{figure}
\epsscale{0.6}
\plotone{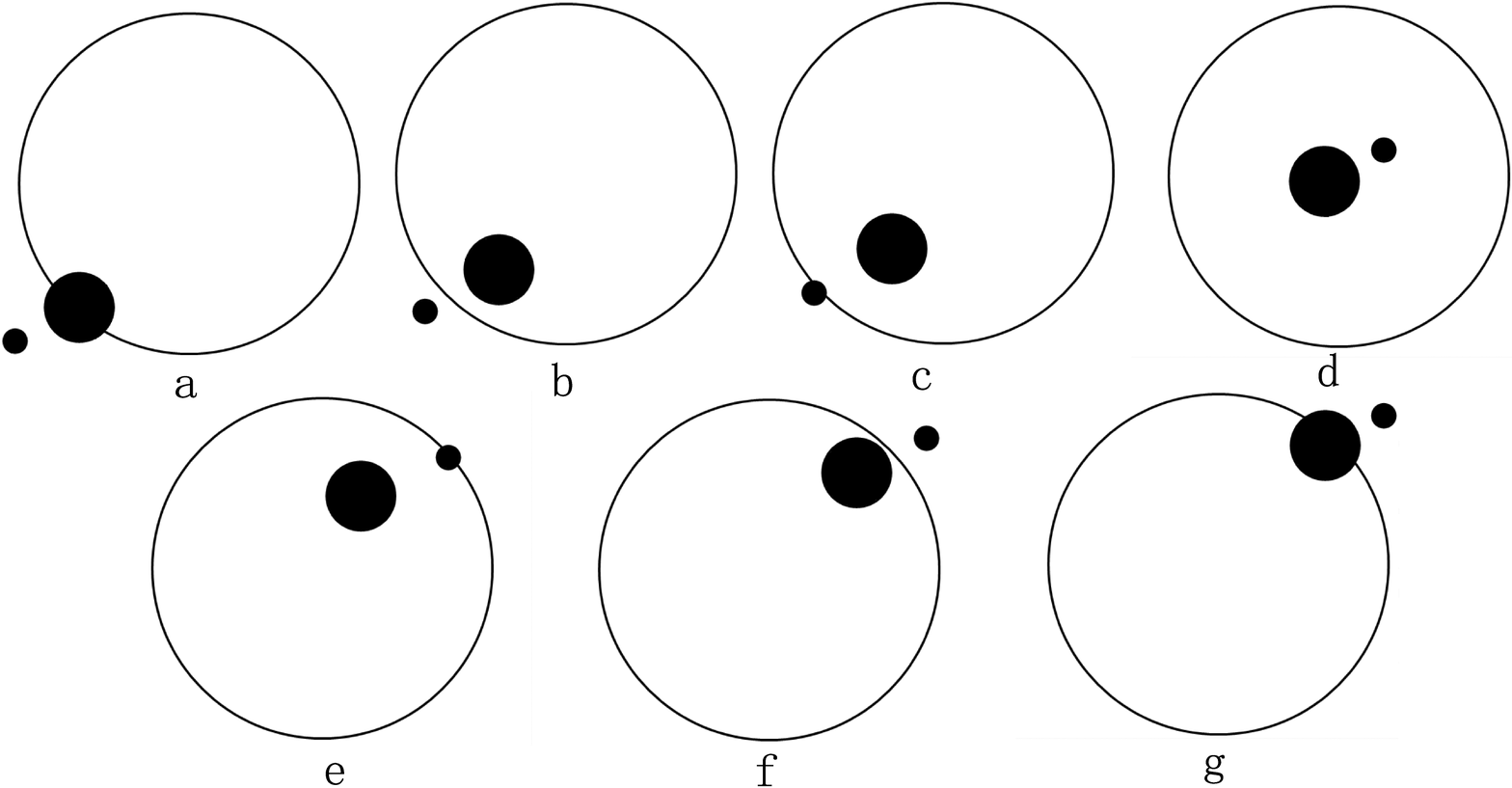}
\epsscale{0.3}
\plotone{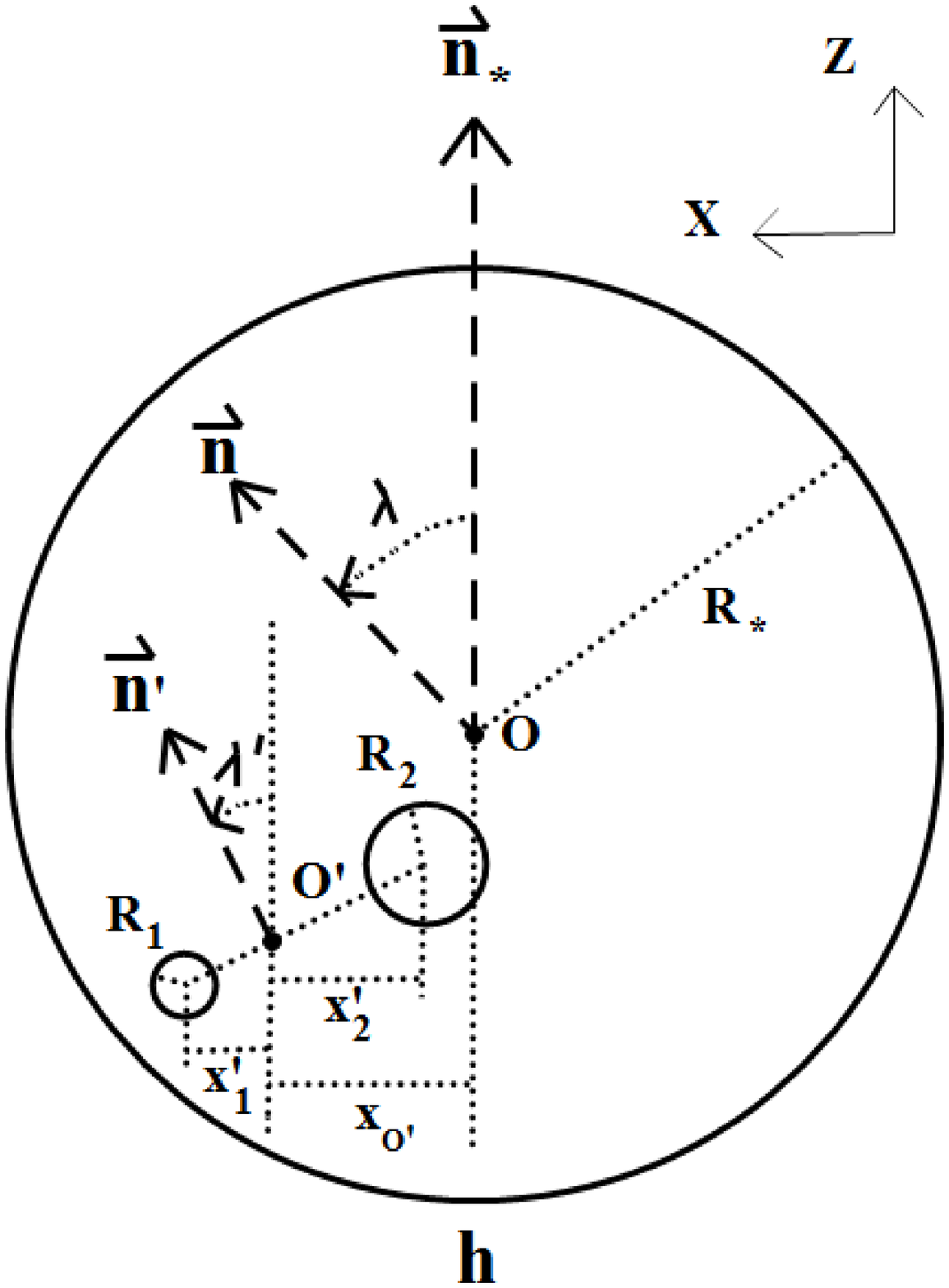}
\caption{Panels (a)--(g): schematic diagram for transit of a binary planet (small shaded circles)
in front of a star (big open circle).
The observer is located in the direction of pointing out of the paper surface.
The binary planet is transiting from left to right.  Panels (a)--(c):
ingress phase; panel (d): complete transit phase of the binary planet; panels
(e)--(g): egress phase. In this example, the rotation of the binary planet
is fast enough so that the small component has changed its relative position from
the left side to the right side of the big one during the complete transit phase (d).
See details in Section~\ref{subsec:transit}. 
Panel (h): schematic diagram for the geometry of the system during the complete
transit phase of a binary planet. The sizes of the star and the binary planet
are all illustrated in the panel. The dashed lines represent the
projection of the labeled vectors onto the sky. 
See also Table~\ref{tab:para} for the meaning of labeled parameters.
}
\label{fig:transit}
\end{figure}

\begin{figure}
\epsscale{1.0}
\plotone{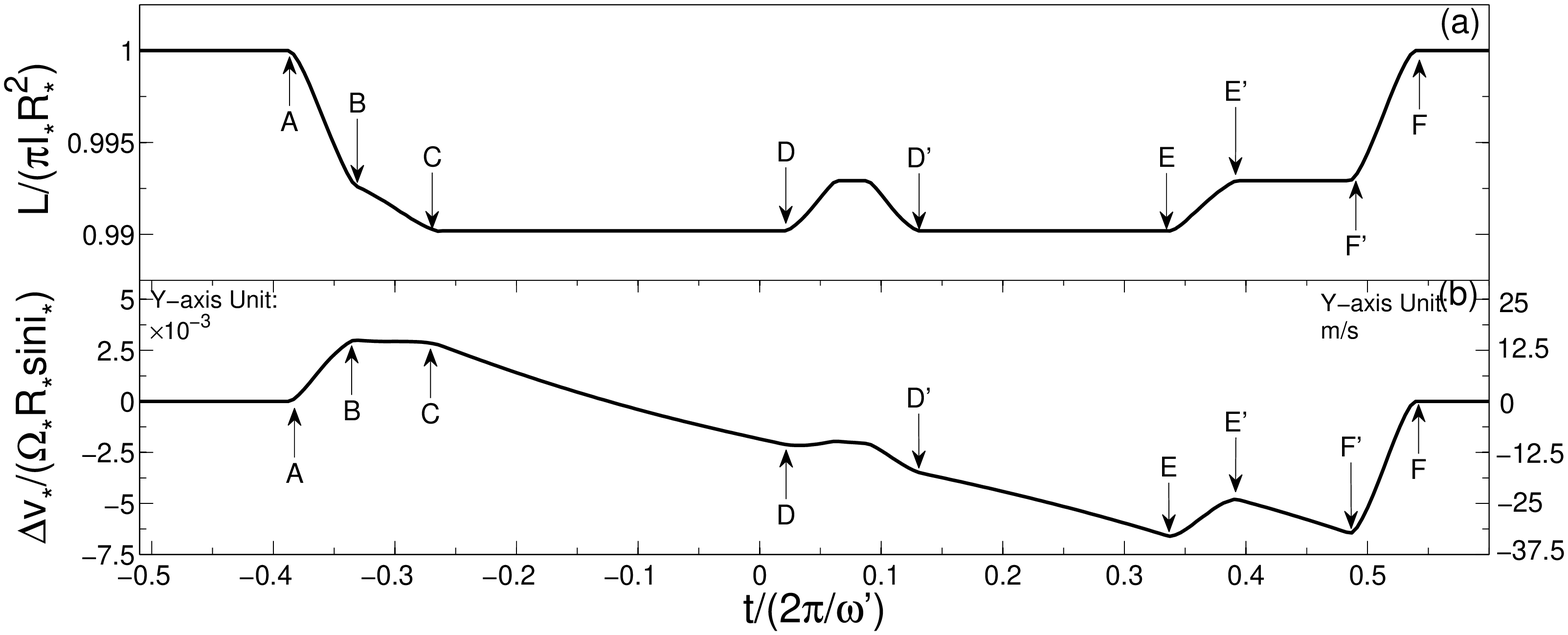}
\caption{Example of transit light curve and stellar radial velocity anomaly for
transit of a binary planet in front of a star. The reflected light from
the surface of the binary planet is ignored for simplicity. Panel (a):
normalized observational stellar light curve $L(t)/(\pi I_*R_*^2)$. Panel (b):
observational radial velocity anomaly of the star. The left y-axis in panel (b)
represents the dimensionless velocity values, and the right y-axis represents
the physical values of the $\Delta v_*$ by assuming
$\Omega_*R_*\sin i_*=5\kms$ (similarly for $\Delta v_b$ in
Fig.~\ref{fig:distribution} and $\Delta v_*$ in Fig.~\ref{fig:triplecurves}.
Related parameters used for this example are as follows:
$\log(\frac{m_1+m_2}{m_*})=-2$, $\rho_1/\rho_2=5$, $m_1/m_2=1.2$,
$a/d=125$, $R_*/a=0.02$, $\frac{A_1+A_2}{A_*}=0.01$, $\lambda=35\degr$,
$i=89.5\degr$, and $\theta=10\degr$.
See details in Section~\ref{subsec:velocity}.}
\label{fig:example}
\end{figure}

When a binary planet transits in front of a star (see a schematic diagram in
Fig.~\ref{fig:transit}), the stellar light curves
may be imprinted with signatures of the binary planet.
We illustrate one example of the normalized transit light curve in
Figure~\ref{fig:example} (see panel a). The observational transit curve is obtained by
$L\equiv\int\int I(x,z)dxdz$, where $I(x,z)$ is the observational stellar
surface brightness at position $(x,z)$. During the transit, part of the
stellar surface is blocked by the planets, and we set
\be
I(x,z)=\cases{ I_*, & unblocked region with $x^2+z^2\le R_*^2$, \cr
        0, & otherwise,}
\ee
For simplicity, the limb-darkening
effect of the stellar surface brightness is ignored in this paper. 
We use a full three-body simulation to obtain the dynamical motion of
the system. The dynamical motion of the binary planet relative to the
star determines the shifting of the blocked region during the transit.
The following phases during the transit are illustrated in
Figures~\ref{fig:transit} and \ref{fig:example}.
\begin{itemize}

\item Ingress phase: at the beginning of the transit, at least one component of
the binary planet starts to block the stellar light, but the projected areas of
the two components have not been fully enclosed by the projected stellar
surface (e.g., the ``AC'' part in Fig.~\ref{fig:example}).

\item Complete transit phase of the binary planet: both the projected areas of
the two components have been fully enclosed by the projected stellar surface
(e.g., the ``CE'' part in Fig.~\ref{fig:example}).
Generally the two planets are more likely to spend most of the transit time in
that phase when $i$ is close to $\pi/2$.
The $d<2R_*$ can be roughly
taken as a condition for the occurrence of this phase during the transit.
If the binary planet has a
sufficiently large $\omega'$, it is likely that during the transit, their
orbital evolution leads to the evolution of their projected areas from
non-overlap to overlap, and to non-overlap again, for which a ``bulge'' (the ``DD$'$''
part in Fig.~\ref{fig:example}) are shown in the light curve.

\item Egress phase: at least one component of the binary planet has transited
to the other end of the projected stellar surface and its projected area is not
fully enclosed by the projected stellar surface again (e.g., the ``EF'' part in
Fig.~\ref{fig:example}).
The ``E$'$F$'$'' flat part
of the light curve in Fig.~\ref{fig:example}a represents the period in which
one component has fully moved out of the projected stellar surface, but the other
one is still completely inside.
\end{itemize}
The transit of a binary is different from that of a single body. As illustrated
above, the binary may enter or exit the transit one by one, or the
projected areas of the two components may overlap, so that some special
features can be shown in the transit light curve (e.g., some step changes or
``bulges'').  Some properties of the binary planet (e.g., radii of its two
components, and its angular velocity and semimajor axis if its $\omega'$ is
sufficiently fast; \citealt{SA09}) can be extracted from the features.  In
addition, the transit duration variation and the transit timing variation
measured from the light curves have also been proposed to obtain the exomoon
mass and the semimajor axis of the moon's orbit \citep{K09a,K09b}. Below we
illustrate that the orbital configuration of the binary planet can be further
constrained by the evolution curve of the observational stellar radial
velocity. 

Note that the velocity in Figure~\ref{fig:example} (similarly in
Fig.~\ref{fig:distribution} below) is expressed in a dimensionless quantity,
where the stellar parameters involved in the normalization could be non-trivial
to estimate in realistic systems and would be done through other independent
methods and abundant knowledge in stellar astrophysics.

\subsection{Stellar radial velocity anomaly and orbital configuration
of a binary planet} \label{subsec:velocity}

As mentioned in the Introduction, a binary planet may leave signature on the
observational radial velocity of the star, as well as on its light curve. The
apparent stellar radial velocity anomaly due to the blocking of the stellar
light is given by
\be
\Delta v_*=-K\frac{\int\int xI(x,z)dxdz}{\int\int I(x,z)dxdz}
\label{eq:Deltav}
\ee
(e.g., \citealt{OTS05,Winn05}), where $K\equiv \Omega_*\sin i_*$ is the line-of-sight
component of the spin angular velocity and may be observationally constrained
from the stellar spectrum. Applying equation
(\ref{eq:Deltav}) to the example shown in Figure~\ref{fig:example}a, we obtain
the evolution curve of
its corresponding radial velocity anomaly in Figure~\ref{fig:example}b (see also
Fig.~1 in \citealt{Simon09}).

Below we demonstrate how the dynamical configuration of the binary planet is
incorporated into the evolution curve of the stellar radial velocity anomaly.
For simplicity, we consider the complete transit phase of the binary planet.
And we analyze the case in which the projected areas of the two planets do not
overlap, and the non-overlap is likely to be true during most of the transit
time especially if $R_1+R_2\ll d$ or $|\pi/2-i'|\ga (R_1+R_2)/d$.
Thus, the stellar radial velocity anomaly can be simplified as follows:
\be
\Delta v_*=K\frac{x_1A_1+x_2A_2}{A_*-A_1-A_2},
\label{eq:Deltavc}
\ee
where $x_1=x_{O'}+x_1'$ and $x_2=x_{O'}+x_2'$ are the $x$-coordinates of the center
of each planet, respectively (see Fig.~\ref{fig:transit}h), 
\be
x_{O'}=a(1-\sin^2\lambda\sin^2 i)^{1/2}\sin[\omega (t-t_0)] \quad (\mbox{if } e=0)
\label{eq:xc}
\ee
is the $x$-coordinate of the center of mass of the binary planet,
$t_0$ is set so that $x_{O'}=0$ at $t=t_0$,
\be
x_1'=d_1(1-\sin^2\lambda'\sin^2 i')^{1/2}\sin[\omega'(t-t_0)+\phi]
\label{eq:x1}
\ee
and
\be
x_2'=-d_2(1-\sin^2\lambda'\sin^2 i')^{1/2}\sin[\omega'(t-t_0)+\phi]
\label{eq:x2}
\ee
are the $x$-coordinates of the two planets relative to their center of mass,
$d_1$ and $d_2$ are the distances of the two planets to their center of mass,
and $\phi$ represents the orbital phase of the inner binary.
The $y$-coordinates of the planets along the line of sight do
not appear explicitly in equation (\ref{eq:Deltavc}), but they are involved
in the expression through the angles $\omega'(t-t_0)$ and $\phi$.
Applying equations (\ref{eq:xc})--(\ref{eq:x2}) to equation (\ref{eq:Deltavc}),
we have
\be
\Delta v_*=\Delta v_{O'}+\Delta v_{b},
\label{eq:Deltav1}
\ee
where
\begin{eqnarray}
\Delta v_{O'} &= & K\frac{a(A_1+A_2)}{A_*-A_1-A_2}(1-\sin^2\lambda\sin^2 i)^{1/2}\sin[\omega(t-t_0)], \label{eq:DeltavO1}\\
\Delta v_b &=& K\frac{d_1A_1-d_2A_2}{A_*-A_1-A_2}(1-\sin^2\lambda'\sin^2 i')^{1/2}\sin[\omega'(t-t_0)+\phi].
\label{eq:Deltavb}
\end{eqnarray}
The $\Delta v_*$ is composed of the two terms, $\Delta v_{O'}$ and $\Delta v_{b}$.
\begin{itemize}
\item The $\Delta v_{O'}$ gives the
radial velocity anomaly
as if the transiting body is a single body with the values of its mass and
projecting area being the total ones of the binary
and contains the orbital configuration of the outer
binary, i.e., the angles $\lambda$ and $i$. This R-M effect due to the transit
of a single planet has been used to extract those angles of some realistic
exoplanetary systems. Together with some
assumption or observational evidence on the possible distribution of the 
inclination of the stellar spin $i_*$, the orbital inclination of the planet
relative to the stellar spin (i.e., the angle between $\vec{n}$ and $\vec{n}_*$) can
be further statistically constrained (e.g., \citealt{Winn09}). This inclination
has also been measured in a number of realistic systems through the photometric anomalies exhibited in transit light curves, which are interpreted as passages
of the planet over dark starspots (e.g., \citealt{SW11}). 

\item The $\Delta v_b$ comes from the relative motion of the inner binary, for which
the sine mode of equation (\ref{eq:Deltavb}) represents its periodical orbital motion. 
The information on the orbital configuration of the inner binary ($\lambda'$, $i'$)
comes from $\Delta v_b$, and we focus on the effects of this term in this paper.
\end{itemize}

Within one transit duration, as $\omega \delta t\sim R_*/a \ll 1$, the
$\Delta v_{O'}$ in equation (\ref{eq:DeltavO1}) can be simplified to be linear
with time as follows,
\be
\Delta v_{O'}=K\frac{a(A_1+A_2)}{A_*-A_1-A_2}(1-\sin^2\lambda\sin^2 i)^{1/2} \omega (t-t_0).
\label{eq:DeltavO2}
\ee
If the sine mode in $\Delta v_b$ can be identified in the observations, its
period and amplitude can be used to constrain the value of $\omega'$ and the
geometric configuration ($\lambda'$, $i'$).  If $\omega'\delta t \sim
\frac{\omega'}{\omega}\frac{R_*}{a}\ll 1$, the $\Delta v_b$ in equation
(\ref{eq:Deltavb}) can be reduced to be also linear with time as follows
\begin{eqnarray}
\Delta v_b &=& K\frac{d_1A_1-d_2A_2}{A_*-A_1-A_2}(1-\sin^2\lambda'\sin^2 i')^{1/2
}[\sin\phi+\omega'(t-t_0)\cos\phi],
\label{eq:Deltavblinear}
\end{eqnarray}
and the slope of the
$\Delta v_*$--$t$ curve during the complete transit phase is given by
\be
k\equiv\frac{d\Delta v_*}{dt}=k_{O'}+k_b\cos\phi,
\ee
where
\be
k_{O'}=K\frac{a(A_1+A_2)}{A_*-A_1-A_2}(1-\sin^2\lambda\sin^2 i)^{1/2}\omega
\label{eq:kOprime}
\ee
and
\be
k_b=K\left(\frac{d_1A_1-d_2A_2}{A_*-A_1-A_2}\right)(1-\sin^2\lambda'\sin^2 i')^{1/2}\omega'.
\label{eq:kb}
\ee
The $k_{O'}$ is constant with time as $\lambda$ and $i$.
The $k_b$ depends on the orbital configuration of the inner binary ($\lambda'$,
$i'$), which is usually constant within one transit duration, as mentioned
in Section~\ref{sec:geometry}. 

Even if the eccentricity of the outer binary $e$ is non-zero, but has a low or
moderate value [e.g. $e$=0.3, so that the linear approximation of
$\omega(t-t_0)$ in eq.~\ref{eq:DeltavO2} is still valid], the expressions for
$\Delta v_{O'}$ and $k_{O'}$ can be modified simply by replacing $a$ with
$ga$ in equations (\ref{eq:DeltavO2}) and (\ref{eq:kOprime}),
where the factor 
\be
g=(1-e^2)^{-1/2}[1+e\cos(\alpha+\angle POQ)],
\label{eq:g}
\ee
the angle $\alpha$ is defined by $\tan\alpha=\sin\lambda\cos
i/\cos\lambda$ and $\cos\alpha=-\cos\lambda/(1-\sin^2\lambda\sin^2 i)^{1/2}$, and
the meaning of the angle $\angle POQ$ is indicated in Figure~\ref{fig:geom}.

\begin{figure}
\epsscale{1.0}
\plotone{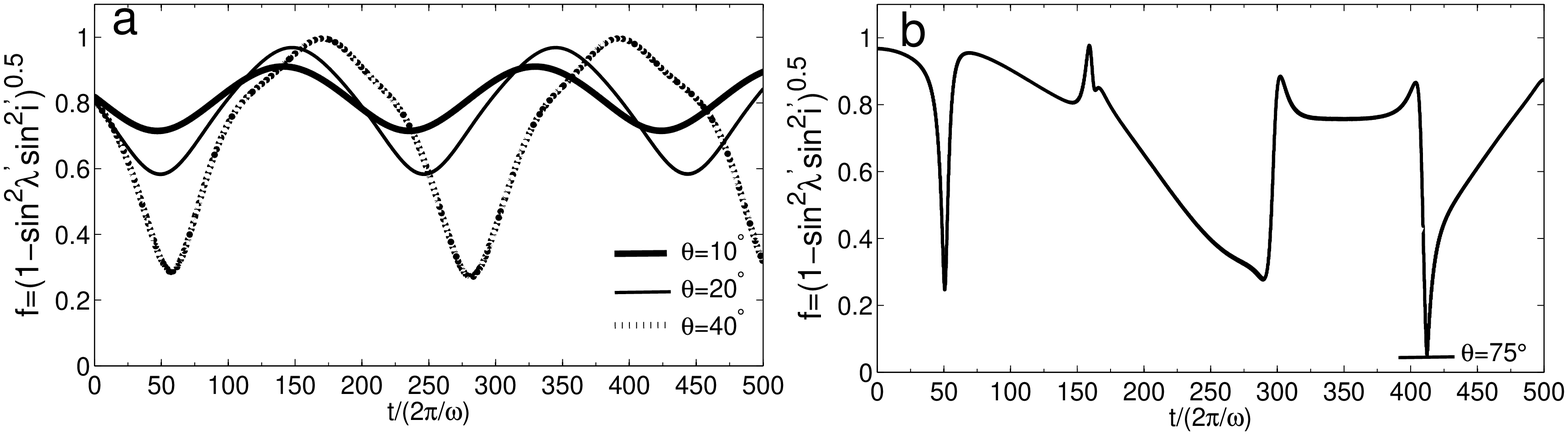}\\
\plotone{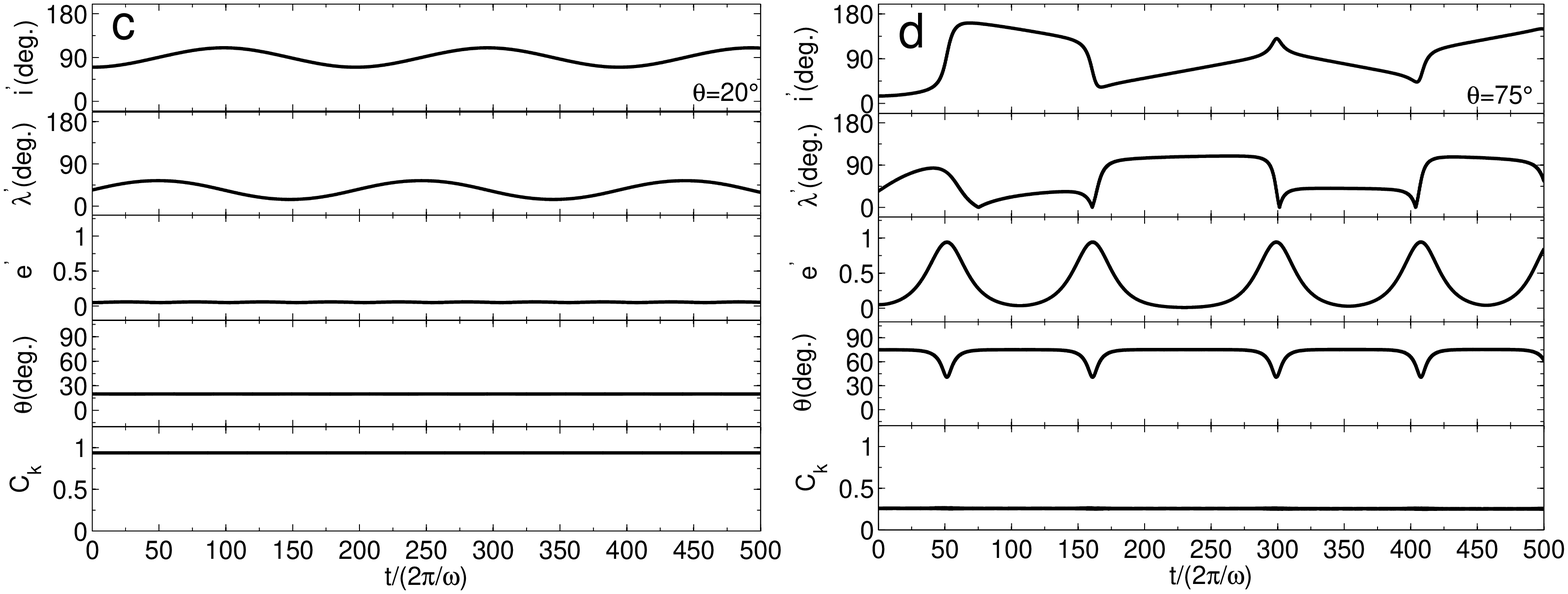}
\caption{Examples of the orbital evolution of a binary planet rotating around a
star, calculated from a full three-body numerical simulation. The related
parameters used are the same as those in Fig.~\ref{fig:example} except the
value of $\theta$. The variable of $f(\theta,t)$ affects the amplitude or slope
of the stellar radial velocity anomaly curve (see eqs.~\ref{eq:Deltavb}
and \ref{eq:kb}). Panel (a) is mainly
for $\theta\la 40\degr$; and panel (b) gives an example with $\theta=75\degr$,
where the Kozai mechanism is effective.  In panel (a), different curves of $f$
represent the evolution under different $\theta$, which produces different
slope distributions of the stellar radial velocity anomaly curves, given
multiple transit observations.  Panels (c) and (d) show the evolution of some
other parameters corresponding to the cases shown in panel (a) and (b),
respectively. In panel (b) ($\theta=20\degr$), the parameters $e'$, $\theta$,
and $C_K$ are almost constant with time, and the angles $\lambda'$ and $i'$
evolve periodically due to the precession of the orbital angular momentum.  In
panel (d), the eccentricity $e'$ and the angle $\theta$ exchange periodically,
while $C_K$ remains constant.  
See details in Sections~\ref{sec:geometry} and \ref{subsec:velocity}.  }
\label{fig:ftheta}\end{figure}

\begin{figure}
\epsscale{0.66}
\plottwo{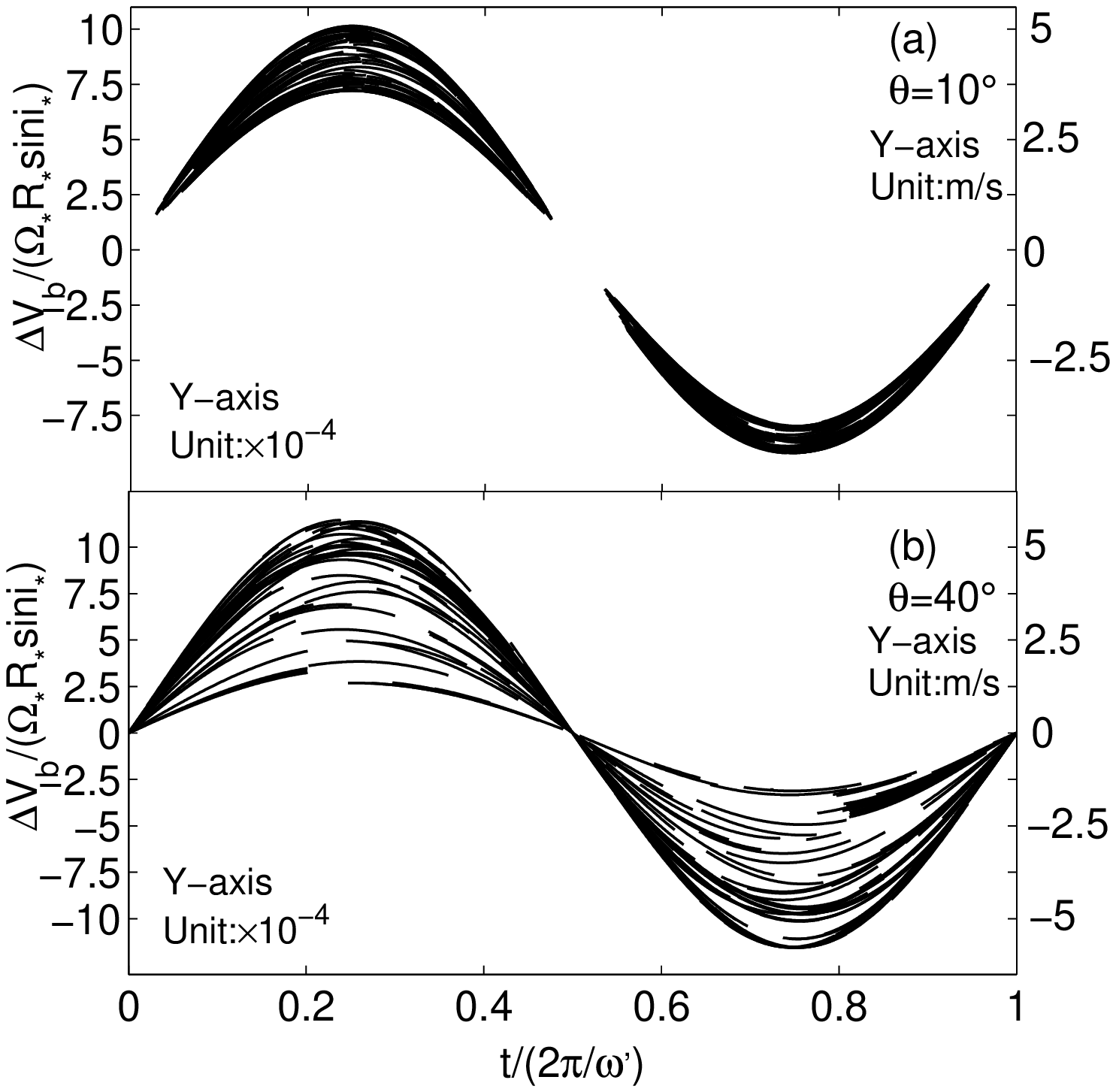}{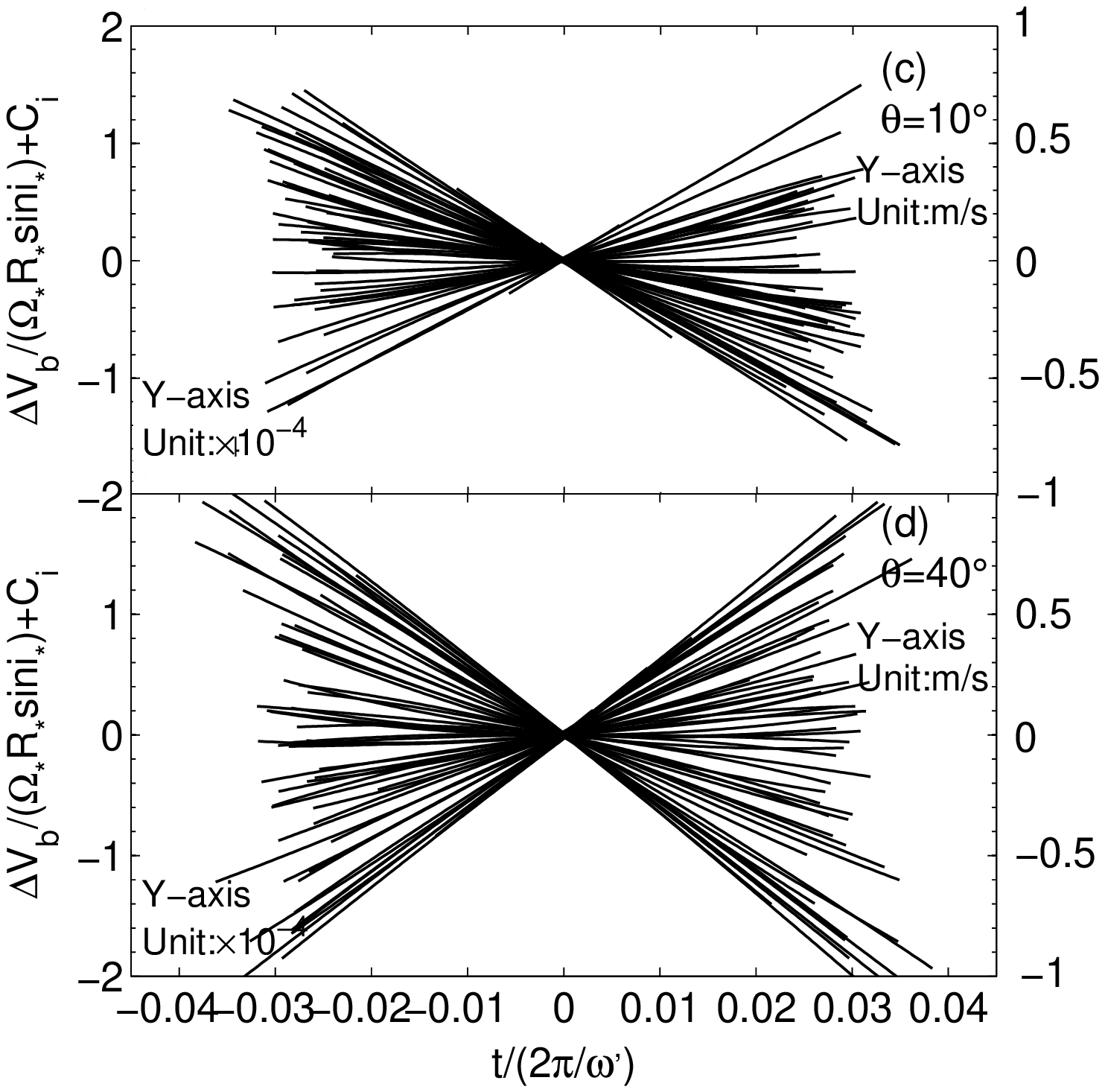}
\epsscale{0.33}
\plotone{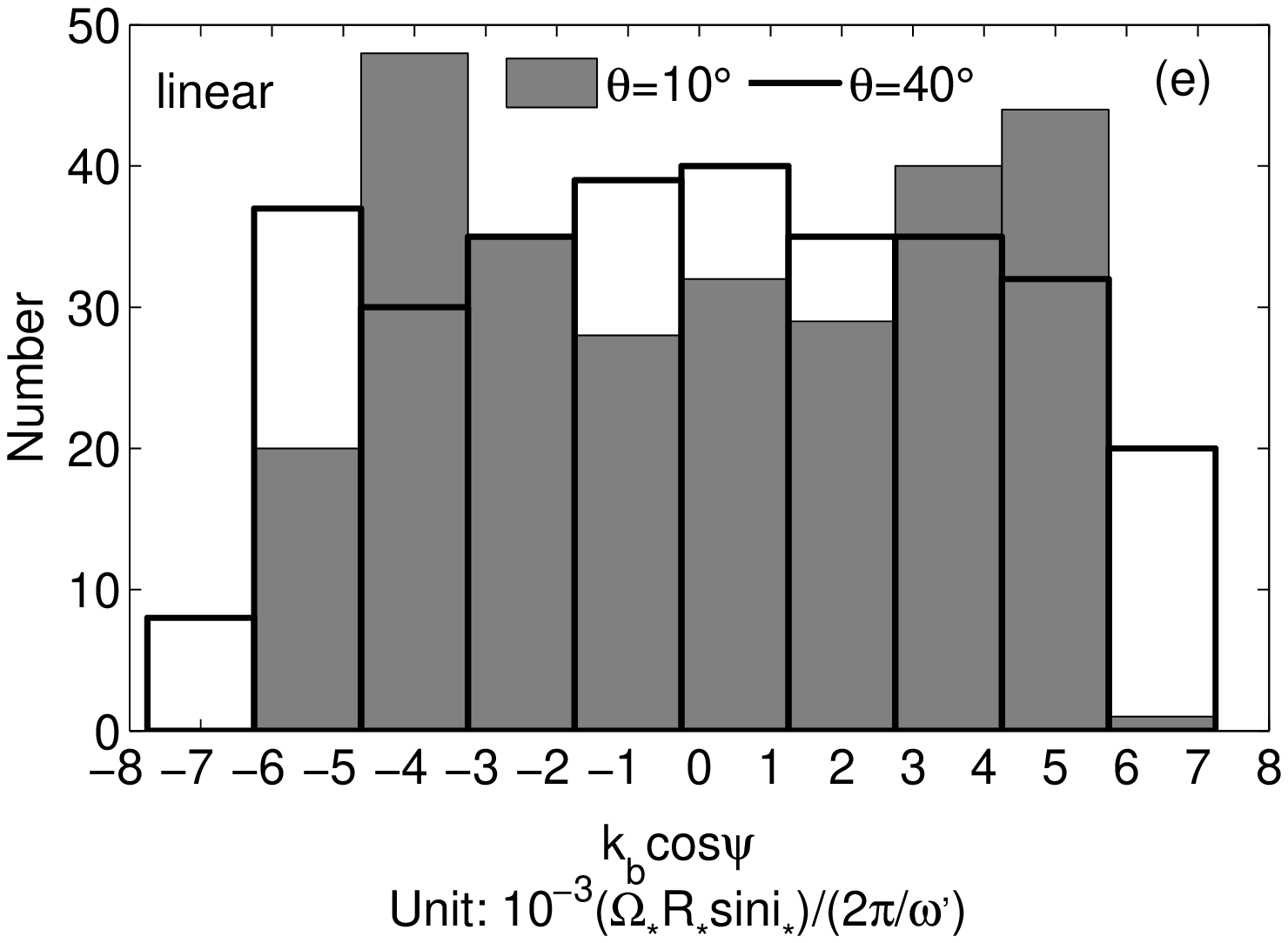}
\caption{The $\Delta v_b$ of the deviation in the R-M effect due to the transit
of a binary planet (see eq.~\ref{eq:Deltavb}). Panels (a)--(b) are for the sine
mode of $\Delta v_b$, and panels (c)--(d) for the linear mode. For each mode,
we show the results for different $\theta$. The figure is obtained by full
three-body numerical
simulations. The related parameters used in the simulations are the same as
those in Fig.~\ref{fig:example}, except $\log(\frac{m_1+m_2}{m_*})=-4$ set for
the linear mode and the different values of $\theta$.  In each of panels
(a)--(d), each curve represents the $\Delta v_b$ within one transit event of
the binary planet, and multiple curves represent the results of its multiple
transit events. We obtain 500 continuous transit events for each system and
display only 20\% of them (chosen randomly) in those panels for visual clarity.
Note that the time span of the transit events is longer than the precession
period $2\pi/\Omega$, and a longer time span can give a better statistical
illustration in panel (e).
Each curve of the transits is different due to the time evolution of the angles
$\lambda'$ and $i'$ and/or different orbital phases $\phi$. All the curves in
each panel are stacked up for comparison of either their amplitudes or their
slopes. In panels (a) and (b), the curves have been shifted along the $x$-axis
to have $\Delta v_b=0$ at $t=0$ and have their corresponding maxima/minima
aligned; and in panels (c) and (d), the curves have been shifted along the
$y$-axis to have their middle points at the zero point, and the constant
$C_i$ on the $y$-axis labels indicates the shifts, which are different
for different events. In each curve, only the complete transit phase is shown
with the overlapping period of the two planets being removed for visual
clarity, which can be done based on its corresponding transit light curve,
although in reality some system information can also be extracted from the
ingress, egress, and overlapping parts.  The projection of a binary planet is
more likely to overlap along the line of sight when $\theta$ is small. The
gap in the middle of the curves shown in panel (a) is just the result of
removing the overlapping phase of the two planets.  As seen from panels (a) and
(b), the scatter of the velocity amplitude in the sine modes is larger for the
larger angle $\theta=40\degr$ than that for $\theta=10\degr$, which is
consistent with the behavior of $f$ shown in Figure~\ref{fig:ftheta}.  Panel
(e) gives the slope distributions of the $\Delta v_b$--$t$ curves obtained
for the linear mode, which also appear different for different $\theta$.  }
\label{fig:distribution}\end{figure}

In a longer time period $\Delta t$($>\delta t$), the evolution of $\lambda'$
and $i'$ is determined by the precession of $\vec{n}$ around $\vec{n}'$
and the value of the angle $\theta$. In this case, we discuss the linear
mode in the following two regimes.
\begin{itemize}
\item If $\Omega \Delta t\ll 1$ (i.e., $\Delta t$ is much shorter than the
precession timescale, but covers multiple transit events), $k_b$ is still roughly constant.
The distribution of the phases $\phi/2\pi$ follows the distribution of
$n\omega'/\omega$ ($n$: integer).  In general, unless $\omega'/\omega$ is an integer, $\phi$
is uniformly distributed between 0 and $2\pi$, and we can get the average
of the slopes over the multiple transits as follows
\be
\overline{k}=k_{O'}, 
\ee
which can be used to constrain the sky-projected angle $\lambda$. The rms of the slopes is
\be
\left[\overline{(k-\overline{k})^2}\right]^{1/2}=\frac{1}{2}k_b.
\label{eq:krms}
\ee
The value of $\theta$ is a function of $\lambda'$ and $i'$ (see
eq.~\ref{eq:theta}), and equation (\ref{eq:krms}) can be used as an
observational constraint for statistical determination of the probability
distribution of $\theta$.

\item If $\Omega \Delta t \ga 1$, $k_b$ changes due to the precession of
$\vec{n}$ and the variation of $\lambda'$ and $i'$.  Here we only discuss the
case with $0\la\theta\la 40\degr$ or $140\degr\la\theta\la 180\degr$.  As seen
from Figure~\ref{fig:ftheta}, the evolution pattern of
$f\equiv(1-\sin^2\lambda'\sin^2 i')^{1/2}$ is different with different
$\theta$.  Note that the evolution pattern for $\pi-\theta$ ($0\la\theta\la
40\degr$) can be obtained by reversing the time in the pattern for $\theta$, as
the precession direction for $\theta>90\degr$ is along $-\vec{n}$. 
The slope of $k$ should distribute within the envelope of $f$, and the
magnitude of the variation of $f$ increases with increasing $\theta$ (for
$\theta<90\degr$). Hence the distribution of the slopes can be used to
constrain the value of $\theta$ (or $|90\degr-\theta|$). We illustrate such an
example in Figure~\ref{fig:distribution}(c)--(e), by doing full three-body
numerical simulations on
dynamical evolution of a binary planet rotating around a star and obtaining its
multiple transiting events.  \end{itemize}

Similarly, if the stellar radial velocity anomaly due to a binary planet
$\Delta v_b$ is in a sine mode in equation (\ref{eq:Deltavb}), observations of
multiple transit events within a longer time $\Delta t$ can be useful to obtain
the evolution of the amplitude of the sine mode, the evolution of the orbital
configuration of the inner binary ($\lambda'$, $i'$), and also further
statistically constrain the angle $\theta$ (see the example shown in
Fig.~\ref{fig:distribution}a--b).

In addition, in some cases, the binary planet are not in the complete transit
phase, but only one component is in the transit. For example, this may occur in
the ingress or egress phase; and if $d>2R_*$, it is also more likely that each
component transits in front of the star one by one. 
In these cases, it is easy to generalize the analysis
above and obtain the deviation in the R-M effect due to the transit of each
component, by setting the projected area of the other component to zero in the
formula above.

Note that in the modeling of most of the other exomoon detection methods, the
orbit of the exomoon surrounding the planet is assumed to be co-aligned with
the orbit of the planet surrounding the star. \citet{SA10} consider how the
special step or overlap features shown in transit light curves (cf.,
Fig.~\ref{fig:example}a) can be used to infer the orbital inclination of an
exomoon, in which the relevant cases are for the condition that $i'$ is
close to $90\degr$. The transit duration variations derived by \citet{K09b}
involve the different inclination parameters of an exomoon, but which is
limited to $\omega'\delta t\ll 1$ or the linear mode and ignores the evolution
of the dynamical systems (i.e., the precession of $\vec{n}'$).

\subsection{Systems likely to be revealed by observations}
\label{subsec:observation}

To have the signatures of a binary planet detectable in observations,
the change of the stellar radial velocity anomaly due to the binary planet
should be significantly large during each transit. As analyzed above, the
deviation $\Delta v_*$ is composed of two parts, $\Delta v_{O'}$ and
$\Delta v_b$ (see eq.~\ref{eq:Deltav1}),
and both have the contributions from a second planet or exomoon.  

For the first part $\Delta v_{O'}$, the contribution from each component of the
binary can be estimated by 
\be
\sim 5\ms \left(\frac{KR_*}{5\kms}\right)\left(\frac{10^3 A_i}{A_*}\right) \quad i=1,2,
\label{eq:DeltavOmoon}
\ee
which is the same if the two components have identical projected areas;
and the special features (e.g., step change EE$'$)
shown in Figure 3 have the same orders of magnitude as that estimate and may
serve as some characteristic signals of binary planet candidates in the R-M
effect.

\begin{figure}
\epsscale{0.7}
\plotone{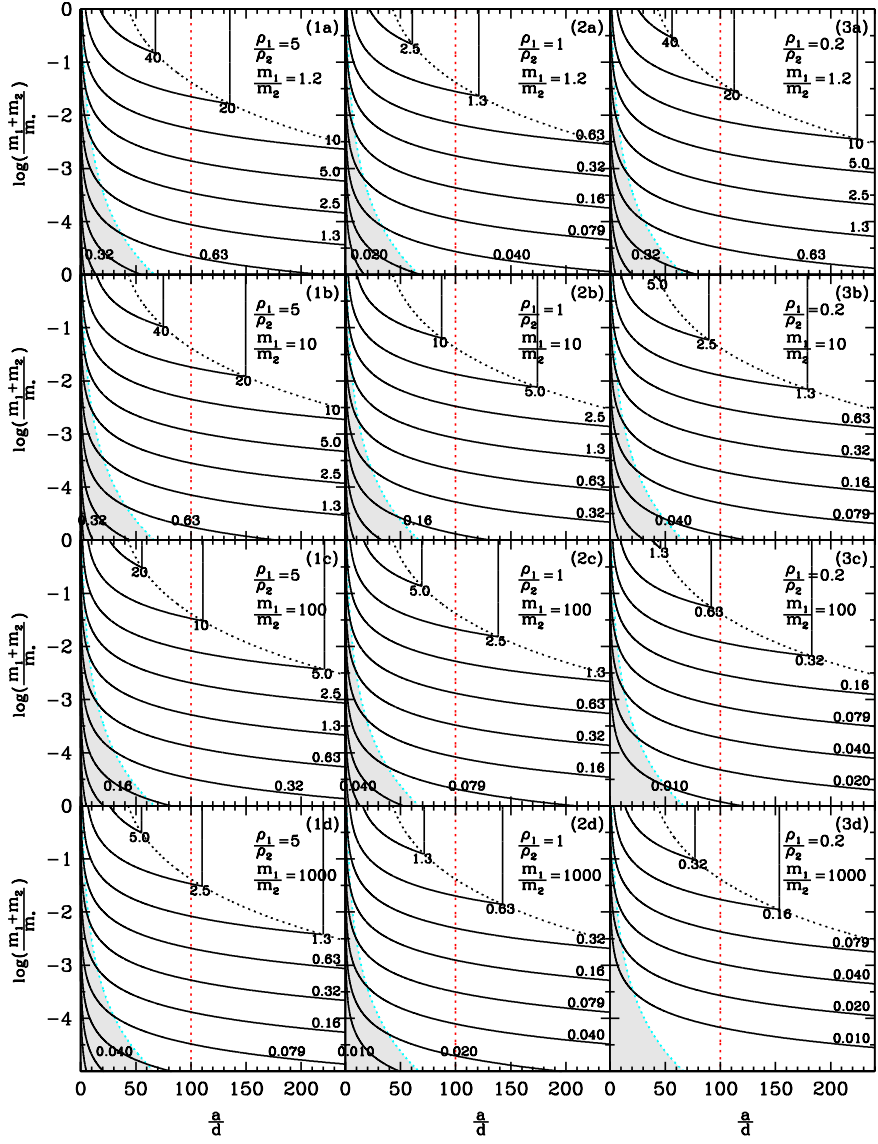}
\caption{Equi-$\delta v_b$ curves for different $\frac{m_1+m_2}{m_*}$ and
$a/d$. The contours of $\delta v_b$ are drawn in black solid curves and
calculated from equations (\ref{eq:deltavb1}) and (\ref{eq:deltavb2}). The
values of $\delta v_b$ labeled for each curve are in units of $\ms$, and they
are chosen in an interval of $\Delta \log(\delta v_b/\ms)=0.3$.  The cyan
dotted curve indicates the Hill radius (eq.~\ref{eq:Hill}), and a binary planet
with parameters located below the curve cannot survive due to the tidal breakup
by the star.  The black dotted curve indicates $\omega'\delta t=1$, which
separates the two regimes expressed by equations (\ref{eq:deltavb1}) and
(\ref{eq:deltavb2}). The region below the black dotted curve has $\omega'\delta
t<1$. The parameters are set as follows: $KR_*=5\kms$,
$\frac{A_1+A_2}{A_*}=0.01$, and $\frac{R_*}{a}=0.005$. The red line is a
reference line for $d=2R_*$. The $\delta v_b$ labeled are estimated for a
binary in the complete transit phase. Generally a binary located to the left of
the red line is more likely not to have the complete transit phase and the
transit of each component in front of the star occurs one by one; in this case,
the $\delta v_b$ can also be easily estimated, and one component may have a
larger value than the labeled one and the other component has a smaller value.
As labeled in the figure, different panels give different mass and mass density
ratios of a binary planet, i.e., $\rho_1/\rho_2=5$, 1, 0.2 and $m_1/m_2=1.2$,
10, 100, 1000.} \label{fig:deltavb1} \end{figure}

\begin{figure}
\epsscale{0.7}
\plotone{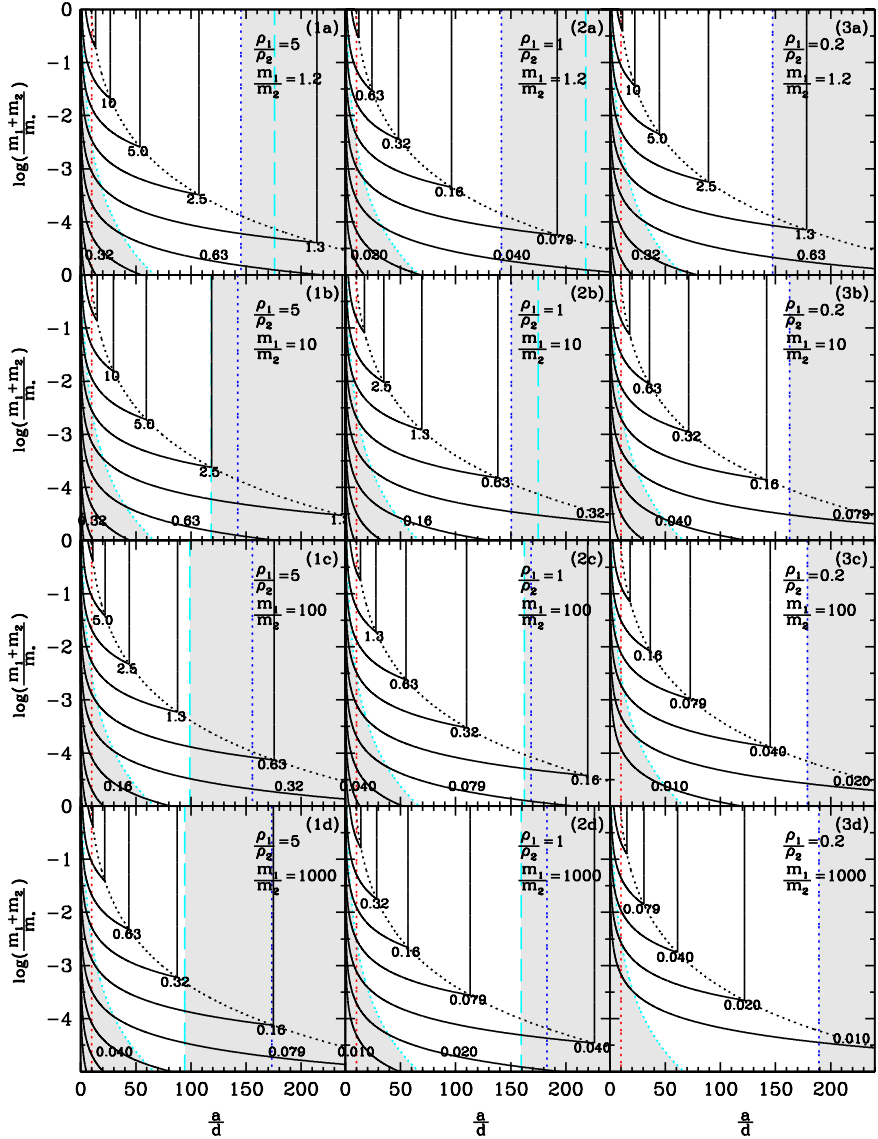}
\caption{Same as in Figure~\ref{fig:deltavb1}, except that the parameter
$\frac{R_*}{a}=0.05$ is set in this figure. The cyan dashed line shown in some
panels represents the Roche limit, calculated from equation (\ref{eq:Roche});
and the parameter space of a binary planet should lie to the left of the line.
The blue dotted line is a reference line for $d=R_1+R_2$, and a binary planet
should also lie to the left of the line.  As in Figure~\ref{fig:deltavb1}, the
grey shaded areas indicate the parameter space that a binary planet cannot
survive dynamically. Both the cyan dashed line and the
blue dotted line are not shown in Figure~\ref{fig:deltavb1}, as they lie beyond
the upper bound of the x-axis of each panel.} \label{fig:deltavb2}\end{figure}

For the second part $\Delta v_b$, which reveals the dynamical configuration of
the inner binary, we used some individual systems with
specific dynamical parameters to indicate its effect in the above section.
Here to see a general parameter space of binary
planet systems that are likely to be revealed in $\Delta v_b$ by observations,
we define the following velocity change:
\begin{eqnarray}
\delta v_b & \equiv & K\left|\frac{d_1A_1-d_2A_2}{A_*-A_1-A_2}\right|\omega'\delta t, \quad \mbox{if } \omega'\delta t<1,
\label{eq:deltavblinear} \\
& \simeq &
KR_*\left(\frac{a}{d}\right)^{1/2}\left(\frac{m_1+m_2}{m_*+m_1+m_2}\right)^{1/2}\left(\frac{A_1+A_2}{A_*}\right)
f_{\delta v_b}
\label{eq:deltavb1}
\end{eqnarray}
and
\begin{eqnarray}
\delta v_b & \equiv & K\left|\frac{d_1A_1-d_2A_2}{A_*-A_1-A_2}\right|, \quad \mbox{if } \omega'\delta t\ge 1,
\label{eq:deltavbsin} \\
& \simeq & 
KR_*\left(\frac{d}{R_*}\right)\left(\frac{A_1+A_2}{A_*}\right)
f_{\delta v_b},
\label{eq:deltavb2}
\end{eqnarray}
where
\begin{eqnarray}
f_{\delta v_b} & = & \left|\frac{\frac{d_1}{d}-\frac{A_2}{A_1+A_2}}{1-\frac{A_1+A_2}{A_*}}\right|, \nonumber\\
&\simeq & \left|\frac{1}{1+m_1/m_2}-\frac{1}{1+\left(\frac{m_1/m_2}{\rho_1/\rho_2}\right)^{2/3}}\right|, \quad \mbox{if } \frac{A_1+A_2}{A_*}\ll 1.
\label{eq:fdeltavb}
\end{eqnarray}
Equation (\ref{eq:deltavblinear}) represents the maximum change of the stellar
radial velocity anomaly due to the inner binary in the linear mode of equation
(\ref{eq:Deltavblinear}), i.e., $k_b\delta t$; and equation
(\ref{eq:deltavbsin}) represents the amplitude of the sine mode shown in
equation (\ref{eq:Deltavb}). Note that the $\delta v_b$ in both equations
(\ref{eq:deltavblinear}) and (\ref{eq:deltavbsin}) is a {\it defined}
variable. Although the expressions are obtained through the limits at
$\omega'\delta t\ll 1$ and $\omega'\delta t\gg 1$, their values at
$\omega'\delta t$ should be in the transition of the two limits and 
the equations above work for the order-of-magnitude estimates and
the purpose of the paper.
As seen from the equations above, not only the planet areas/sizes are involved
in the R-M effect as for a single planet, but their masses or mass densities
are also involved in the effect for a binary planet due to the relative motion
of the two components of the binary, as the hidden areas of the stellar
surface are affected by the relative positions of the binary components and the
relative distance of each component from the center of mass of the binary (cf.,
eqs.~\ref{eq:x1} and \ref{eq:x2}) is determined by its two component masses.
Given the mass ratios of the two components, the radius ratios can also be
expressed through their mass density ratios. Thus, the
amplitudes of the R-M effects indicated in
Figures~\ref{fig:deltavb1}--\ref{fig:triple} are expressed through the extra
dimensions in multiple panels. The involvement of the extra (mass density)
dimension in the study is useful especially considering that recent Kepler
observations have revealed the mass density of planets do span a large range.
In addition, the R-M effect for the transit of a single planet is related with
the orbital semimajor axis, but the semimajor axis $a$ of the outer binary is
involved in the effect for a binary planet as shown in equation
(\ref{eq:deltavb1}) because the transit time $\delta t$ is affected by $a$ and
a longer transit time leads to a larger change of the relative position of the
binary components and further a larger deviation in the stellar radial
velocity.

As seen from equation (\ref{eq:fdeltavb}), the value
of $\delta v_b$ is significant only if the two components of the binary planets
are different, especially if the heavy one has a high mass density and the
light one has a low mass density.  If the two components have the same mass and
mass density, they have the same projecting area and their motion is symmetric,
and thus we have $\delta v_b=0$ with $f_{\delta v_b}=0$, although in this case
the contribution from $\Delta v_{O'}$ (eq.~\ref{eq:DeltavO1}) can be large due
to the combination of the projected areas of the two single components. 
In addition, the value
of $\delta v_b$ is significant only if the area of the binary planet $A_1+A_2$
is significantly large, as $\delta v_b$ is proportional to
$\frac{A_1+A_2}{A_*}$.

By using equations (\ref{eq:deltavb1}) and (\ref{eq:deltavb2}), we show how
$\delta v_b$ depends on other various properties of the system in
Figures~\ref{fig:deltavb1}--\ref{fig:deltavb2}. We set the parameters
$KR_*=5\kms$ and $\frac{A_1+A_2}{A_*}=0.01$ in both of the figures. We set
$R_*/a=0.005$ and $0.05$ in Figures~\ref{fig:deltavb1} and \ref{fig:deltavb2},
respectively. If $R_*=R_\odot$, $R_*/a=0.005$ corresponds to the distance of
the earth from the sun ($a=1\AU$), and $R_*/a=0.05$ corresponds to some typical
semimajor axis of hot Jupiters discovered in the vicinity of a star
($a=0.1\AU$). The contours of $\delta v_b$ as a function of $(m_1+m_2)/m_*$ and
$a/d$ are shown by black solid curves. In each figure we display how the
contours change for binary planets with different mass ratios
($m_1/m_2=1.2,10,100,1000$) and different mass density ratios
($\rho_1/\rho_2=5,1,0.2$). 
As seen from the figures, in the region below the black dotted curve (i.e.,
$\omega'\delta t<1$), $\delta v_b$ increases with increasing $(m_1+m_2)/m_*$
and $a/d$, as $\omega'\delta t$ does in the same tendency (see
eq.~\ref{eq:deltavb1}); and in the region above the curve, $\delta v_b$
increases with decreasing $a/d$ (given $R_*/a$), as the difference of the
line-of-sight rotation velocity of a star covered by each component is likely
to be relatively large for a wide binary planet (with large $d$).  By comparing
the contours obtained for different mass density ratios, the figures also
illustrate that the values of $\delta v_b$ can be significant (e.g., up to
$\ms$ or several ten $\ms$) only if the two components of the binary planets
are different, especially for high $\rho_1/\rho_2$ ratios (see panels 1a-1c),
as analyzed above.  Given $\rho_1/\rho_2$, the $\delta v_b$ at bottom panels
(e.g., 1d, 2d, 3d) have relatively low values, which indicates the difficulty
to detect the effect of a too small exomoon. Figure~\ref{fig:deltavb2} has a
higher $R_*/a$ than Figure~\ref{fig:deltavb1}. On the one hand, the curve of
$\omega'\delta t=1$ in Figure~\ref{fig:deltavb2} (black dotted curve) shifts
downwards; and thus, although the contours of $\delta v_b$ below the black
dotted curve are the same in both of the figures, the region above the curve
has relatively low $\delta v_b$ in Figure~\ref{fig:deltavb2}. On the other
hand, a higher $R_*/a$ would imply a shorter period of the binary rotating
around the star and thus a higher probability to do multiple transit
observations to get a better statistics for the system.  In addition, for
different values of $KR_*$ and $\frac{A_1+A_2}{A_*}$, the contour values of
$\delta v_b$ in the figures should be adjusted simply by multiplying them by a
factor of $(\frac{KR_*}{5\kms})\cdot(\frac{A_1+A_2}{0.01A_*})$.

Based on the results above, we discuss the magnitude of the $\delta v_b$ in the
following examples of binary planet or exomoon systems. We assume that the
central star is a solar-like star (with solar mass and radius).
\begin{itemize}

\item An earth-moon system (with $m_1/m_2\simeq81$, $\rho_1/\rho_2\simeq 1.6$,
$R_*/a\simeq0.005$, $a/d\simeq 400$)
may cause a $\delta v_b$ only up to $0.1\cms$, e.g., due to the small sizes
of the earth and the moon (with $\frac{A_1+A_2}{A_*}\simeq10^{-4}$), which is
too small to be detected. Note that according to the estimate by equation
(\ref{eq:DeltavOmoon}), the contribution of the moon to the
deviation $\Delta v_O'$ can be up to $1\cms$ (consistent with the value shown
in \citealt{Simon10}, where the effect of $\Delta v_b$ is not discussed).

\item The Ganymede is the biggest moon in the Solar system. A Jupiter-Ganymede
system (with $m_1/m_2\simeq 1.3\times10^{4}$, $\rho_1/\rho_2\simeq 0.7$)
may lead to a $\delta v_b$ only up to several $\cms$ (for $R_*/a\simeq 0.001$,
$a/d\simeq 800$).

\item For a Jupiter-rocky moon system (e.g., with $\rho_1/\rho_2=0.2$): a moon
or satellite
with mass $m_2<0.1m_1$ has a low $\delta v_b$ less than $\ms$ (cf., panels
3b-3d in Figs.~\ref{fig:deltavb1}--\ref{fig:deltavb2}). A Jupiter-earth system
(with $m_2\simeq 3\times 10^{-3}m_1$) has a low $\delta v_b$ only in the range
of 1--10$\cms$ (see panels 3c-3d). However, if the rocky moon/satellite has a mass
close to the Jupiter (see panel 3a), the $\delta v_b$ can be up to several to
several ten $\ms$, which may be detectable by the current techniques. Note that
this is an exotic case, as all the rock bodies revealed by the Kepler do not
exceed several ten earth mass so far\footnote{http://kepler.nasa.gov/}.

\item A binary Jupiter-like planet system (e.g., with component masses
$m_1=10M_J$ and $m_2=1$ or $0.1M_J$, and with Jupiter-like mass densities
$\rho_1=\rho_2$) may have a $\delta v_b$ ranging from $\ms$ to several ten
$\ms$, although hot binary Jupiter systems (with relatively high $R_*/a$) have
relatively low $\delta v_b$. The $\delta v_b$ can be even larger if
$\rho_1>\rho_2$.  For such a system, if any, its signatures on the R-M effect
may be detected by future observations.

\item The CoRoT-9 system has a solar-like central star and its orbiting
exoplanet CoRoT-9b has the mass and radius close to the Jupiter's. The CoRoT-9b
is one of the longest period transiting Jupiter ($\simeq 95$ days) that has so
far been confirmed and has a semimajor axis $a\simeq
0.4\AU$\footnote{http://exoplanet.eu/}.  Its $R_*/a\simeq 0.01$ is between the
cases shown in Figs.~\ref{fig:deltavb1} and \ref{fig:deltavb2}. As inferred
from the figures, if the CoRoT-9b has a satellite $m_2\ga 10^{-3}m_1$, its
$\delta v_b$ can range from several $\cms$ to several $\ms$, depending on the
detailed satellite properties.

\end{itemize}

By using equations (\ref{eq:DeltavO2}), (\ref{eq:kOprime}),
(\ref{eq:deltavblinear}), and (\ref{eq:deltavbsin}), 
the ratio of the two parts in the deviation $\Delta v_*$ (see 
eq.~\ref{eq:Deltav1}) is about 
\be
\frac{\delta v_b}{k_{O'}\delta t}\simeq 0.1\left(\frac{\delta
v_b}{5\ms}\right)\left(\frac{5\kms}{KR_*}\right)\left(\frac{0.01A_*}{A_1+A_2}\right).
\ee
As mentioned above, both of the two parts have the contribution from a 
second planet or exomoon, and the ratio of the two contributions can
be estimated by 
$(\frac{\delta v_b}{5\ms})(\frac{5\kms}{KR_*})(\frac{10^{-3}A_*}{A_i})$.

Note that a binary planet has a different gravitational effect on the stellar
motion from
a single planet. The quadrupole moment of the gravitational force from the
binary planet is
\be
\Delta F\sim \frac{Gm_*}{a^2}\cdot\frac{m_1d_1^2+m_2d_2^2}{a^2},
\ee
and its effect on the dynamical motion over each transit duration can be
estimated by
\be
\frac{\Delta F\delta t}{m_*}\sim1\cms\left(\frac{v_*}{100\ms}\right)\left(\frac{1
0d}{a}\right)^2\left(\frac{10m_2}{m_1}\right)\left(\frac{10R_*}{a}\right),
\ee
which is generally negligible. The $v_*$ is the velocity of the star relative to the center of mass of
the system. It would be
interesting to investigate the long-term dynamical effect of the binary planet
on the stellar radial velocity, but which is beyond the scope of this paper.

\subsection{Application to hierarchical triple star systems}\label{subsec:triple}

\begin{figure} 
\epsscale{0.9}
\plottwo{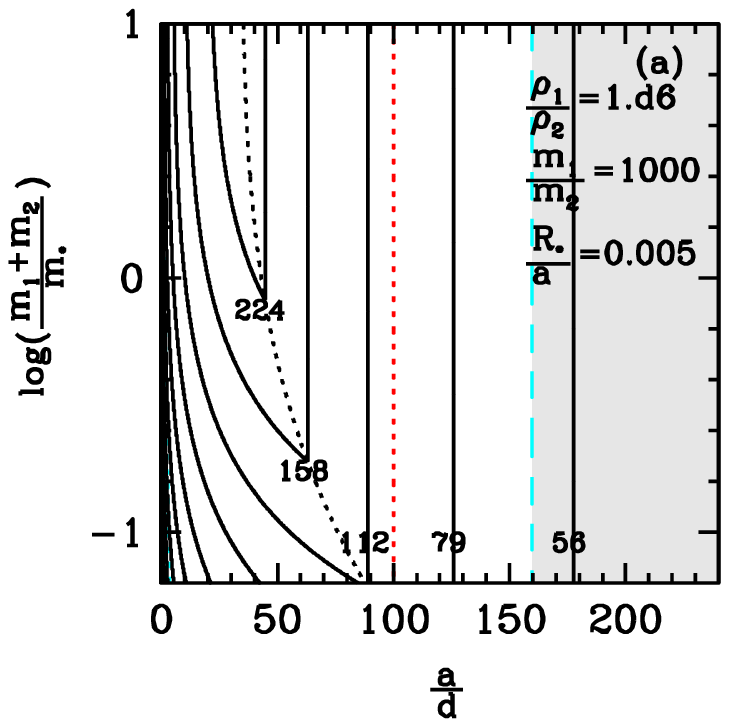}{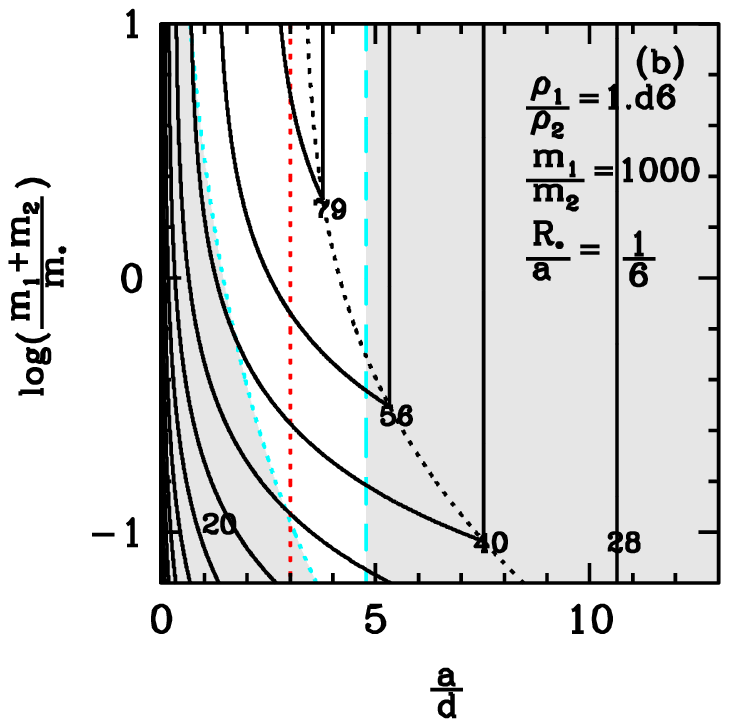}
\caption{Equi-$\delta v_b$ curves for hierarchical triple star systems.
The curves have the same meanings as those in Figures~\ref{fig:deltavb1} and
\ref{fig:deltavb2}. The values of $\delta v_b$ labeled for each curve are 
in an interval of $\Delta \log(\delta v_b/\ms)=0.15$.
The binary star in the hierarchical triple star system
is composed of a compact object (e.g., white dwarf or neutron star) plus
a planet/brown dwarf, so a very high density ratio $\rho_1/\rho_2$ is used 
in the figure. The parameters $KR_*$ and $\frac{A_1+A_2}{A_*}$ are
the same as those in Figure~\ref{fig:deltavb1}. The $R_*/a$ is $0.005$ for
panel (a) and $1/6$ for panel (b), respectively.
As $m_1/m_2$ decreases, the
cyan dashed line shifts rightwards, and the contours of $\delta v_b$ are
affected little for $m_1/m_2\ga 10$. See also Section~\ref{subsec:triple}. 
}
\label{fig:triple}\end{figure}

Figures~\ref{fig:deltavb1}-\ref{fig:deltavb2} can be applied to a hierarchical
triple star system (e.g., for the parameter space $\frac{m_1+m_2}{m_*}\ga
10^{-3}$, in which a dark binary star is transiting in front of a tertiary star
(e.g., \citealt{Carter11}). Here by ``dark'' we mean that the light emission
from the binary is ignored as planets for simplicity. It is easy to generalize
the analysis above to include the light emission from the binary.  If the
transiting binary is a compact object (e.g., white dwarf or neutron star) plus
a planet/brown dwarf, the factor $f_{\delta v_b}$ is up to 1, and the $\delta
v_b$ can be much larger than that of binary planet system.
Figure~\ref{fig:triple} illustrates such a case. As seen from the figure, the
$\delta v_b$ is high, up to $10^2\ms$. Thus, such hierarchical triple star
systems may be revealed through the R-M effects by future observations.

\begin{figure}\epsscale{0.8}
\plotone{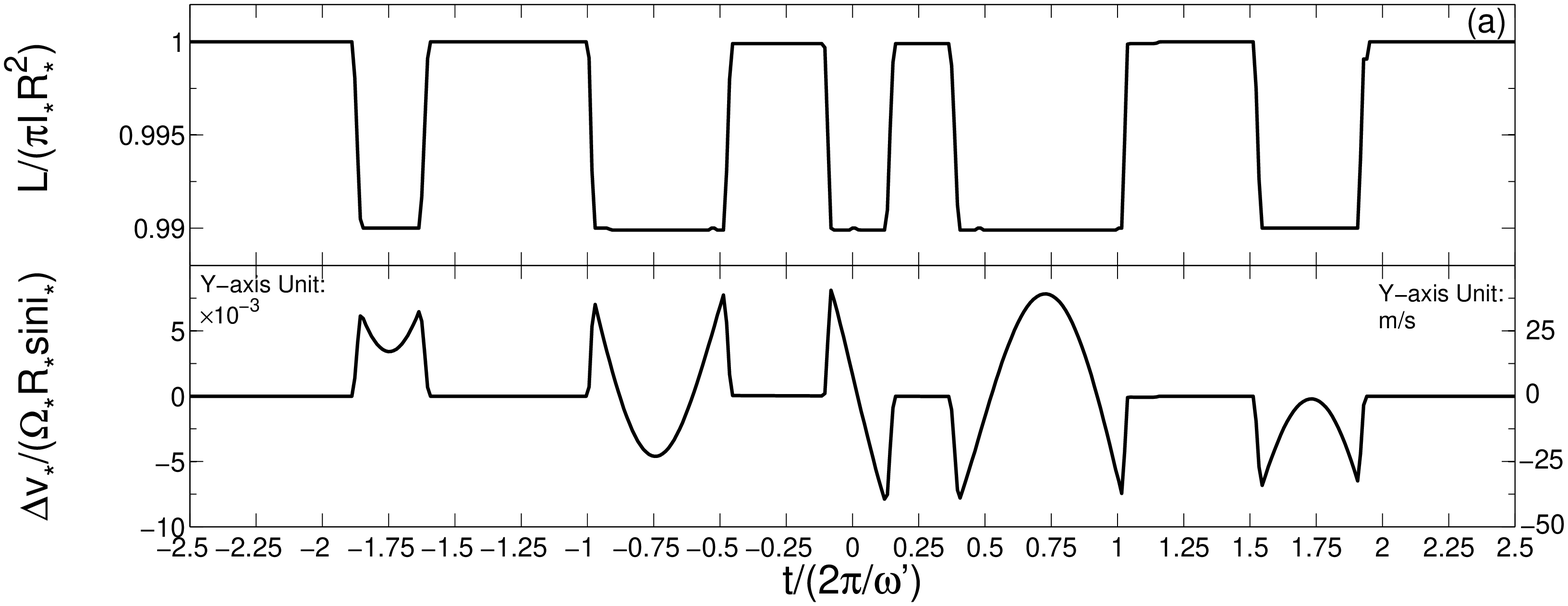}\\
\plotone{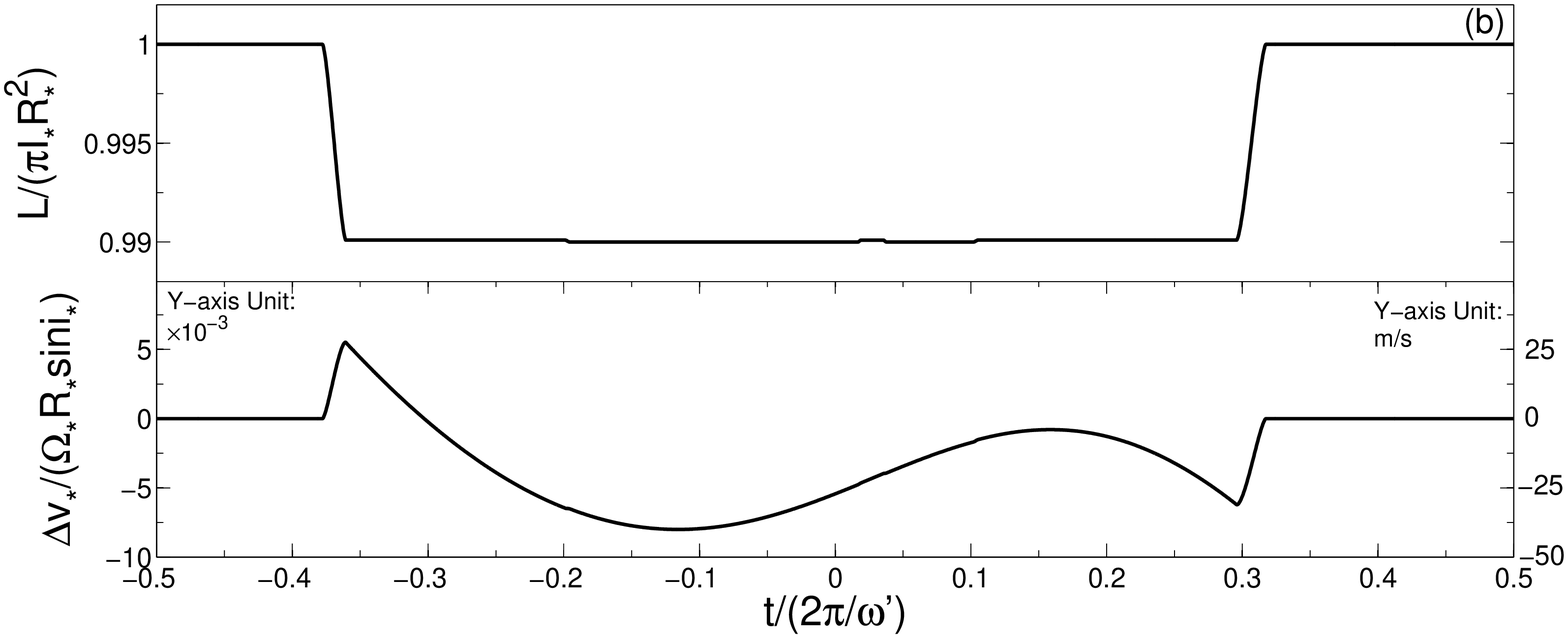}
\caption{Examples of the simulated transit light curves and the stellar radial
velocity anomaly curves for a dark binary star transiting in front of a
tertiary star. The methods to obtain the simulated curves are the same as those
for Figure~\ref{fig:example}.  The system parameters used in panels (a) and (b)
are chosen from the parameter space shown in Figure~\ref{fig:triple}(a) and
(b), respectively. Panel (a) has $\log(\frac{m_1+m_2}{m_*})=0$, $a/d=150$,
$i=90\degr$, and $\theta=10\degr$; and panel (b) has
$\log(\frac{m_1+m_2}{m_*})=0$, $a/d=4$, $i=89.5\degr$, and $\theta=10\degr$.
Both of the examples indicate a significant amplitude of changes in the radial
velocity anomaly caused by the relative motion between the compact object and
the planet/brown dwarf. In panel (a), a number of the bulge/hill/trough
features are indicated in the curves, which is because the orbital period of
the dark binary is relatively short so that the planet/brown dwarf rotating
around the compact star can move completely into, and then move out of, and
then re-move into the projected stellar surface before the complete transit
phase.  The similar effect of the rotational motion also occurs during the
egress phase.  In panel (b), the sine mode of $\Delta v_b$ represented by
equation (\ref{eq:Deltavb}) is significantly shown in $\Delta v_*$.  See also
Section~\ref{subsec:triple}.} \label{fig:triplecurves}\end{figure}

Figure~\ref{fig:triplecurves} illustrates two examples of the transit light
curves and the radial velocity anomaly curves for the triple star systems with
dynamical parameters located in the parameter space shown in
Figure~\ref{fig:triple}.  As seen from the figure, the radial velocity anomaly
caused by the rotational motion of the planet/brown dwarf around the compact
object is much more significantly displayed either through the
bulge/hill/trough features during the ingress/egress phase (for panel a) or in
the sine-like curve during the complete transit phase.  The values of
$\Delta v_b$ can be comparable to $\Delta v_{O'}$. It is plausible to
expect that the features of these curves are very useful to infer the dynamical
configurations of the systems, as illustrated for binary planets above. A more
detailed discussion on extracting the configurations is beyond the scope of the
paper.

\section{Discussion} \label{sec:discussion}

The properties of binary planets (e.g., mass, size, and semimajor axis) can be
constrained through their signatures on the transit light curves and stellar
radial velocity curves. Our analysis of the R-M effect for a binary planet
during its complete transit phase show that effect is composed of two parts.
The first part is the conventional one similar to the R-M effect from the transit
of a single planet with the combined masses and projected areas of the binary
components (eq.~\ref{eq:DeltavO1});
and the second part is caused by the orbital rotation of the binary components,
which may add a sine- or linear-mode deviation to the stellar radial velocity
curve (eq.~\ref{eq:Deltavb}).

In this paper we focus on the discussion on the
second part, and we find both the evolution of the orbital
rotating phases and the precession of the binary orbital plane may lead to
different amplitudes of the deviations in different transit events of the same
system. The resulted distribution and dispersion of the deviations in multiple
transit events can be used to extract the orbital configuration of the binary
planet or even the inclination of its orbital plane relative to the plane of
its center of mass rotating around the star (e.g., Fig.~\ref{fig:distribution}).

The second part of the R-M effect is more likely to be revealed if the binary
components have different masses and mass
densities, especially if the heavy one has a high mass density and the light
one has a low density. For example, our calculations show that the signature
can be up to several or several ten $\ms$ with the mass ratio $m_1/m_2$ up to
$10^{3}$, if the mass density ratio $\rho_1/\rho_2=5$. A small and rocky
exomoon with $m_2<0.1m_1$ would cause a low $\delta v_b$ less than $\ms$. 
A strong signature may be caused if at least one of the components of the
binary planet is a giant planet.
Note that stellar noises produced by oscillations, granulation phenomena, and
activities could contribute to the change of stellar radial velocities with an
amplitude up to $\ms$ (e.g., \citealt{Dumusque11}), but
they have their own variation periods and patterns to be distinguished from the
effect of binary planets, and some statistical methods can be developed to
extract smaller signals from the noises.

A long observation time would cover multiple transits of a binary planet, and a
better statistics on the distribution of the deviation in the R-M effect could
be potentially obtained.  To trace the evolution of the geometric configuration
of the system, we need at least one period of the orbital angular momentum of
the inner binary precessing around that of the outer binary. The precession
period is roughly about $\omega'/\omega$ times the orbital period of the outer
binary. For the parameter space shown in Figures~\ref{fig:deltavb1} and
\ref{fig:deltavb2}, $\omega'/\omega$ is generally less than $200$ (cf., the
$\omega'\delta t=1$ curve, which can also be taken as a reference line for
$\omega'/\omega\simeq (R_*/a)^{-1}$).

The misalignment between the plane of the binary planet and its rotating plane
around the star is one of fundamental parameters of the dynamical system.  A
large misalignment is likely to cause a large dispersion or different
distribution of the deviations in the R-M effect. Recent measurements have
discovered that the orbit of a planet may be highly inclined to the stellar
spin.  Different mechanisms have been proposed for the formation of those
misaligned orbits, e.g., Kozai capture, planet-planet scattering, resonance
capture by planet migration (e.g., \citealt{MCS11,NIB08,FT07,YT01}).
Similarly, the different configurations/inclinations of binary planets
(see $\theta$ defined in Table~\ref{tab:para}), if detected in
future, should be also useful in constraining formation mechanisms of binary
planets or exomoons, and shed new light on our understanding the diversity of
planetary systems.  
For example, different moon formation mechanisms may lead to their different
kinematic distributions. Moons formed from the disk material surrounding a
planet are predicted to have prograde orbits, while those formed from
gravitational capture/impacts/exchange interactions can have either prograde or
retrograde orbits (\citealt{JH07}; and references therein). It is
also likely that the orbit of a moon could be affected by the later evolution
of the system (e.g., by the later inner/outer migration of the outer/inner
planets). Regarding a binary planet system with comparable component masses,
the study of their formation theory is starting (e.g., \citealt{P10}), and one
of the most exciting steps would be to discover a realistic system in
observations in the near future.

To discover binary planets and exomoons becomes promising and practical with
future developments in instruments, which also lay the foundation for finding
the signatures discussed in this paper.  For example, planned ground-based
surveys such as the {\it Large Synoptic Survey Telescope} may detect thousands
to tens of thousands of planetary transit candidates; and the space missions,
{\it PLAnetary Transits and Oscillations of stars} and {\it Transiting
Exoplanet Survey Satellite}, aim to find transiting planets around relatively
bright stars, making it easier to confirm discoveries using follow-up radial
velocity measurements. Hopefully follow-up observations could provide some
binary planet or exomoon candidates.  The
astro-comb technique is aiming to achieve a precision as high as $1\cms$ in
astronomical radial velocity measurements \citep{Li08}.  \citet{Kipping12} also
proposed a systematic search for exomoons around transiting exoplanet
candidates observed by the {\it Kepler} mission.

We studied the basic signatures of binary planets that are likely to
be revealed in the R-M effect. To understand the roles of the crucial
parameters played in the signatures, we have made some approximations
in our study, which could be improved or adapted to realistic systems
in future work, for example, the study could be
extended to a general case in which the center of mass of the binary
planet is on an eccentric orbit.  The study would become complicated if an
exoplanet has multiple moons. In this case, the small moons would contribute
little to the deviation in the R-M effect, and the one with a relatively large
radius and located at a relatively far distance from the primary planet would
imprint the most significant effect.
The limb-darkening effect can reduce the amplitude of the R-M effect by 20-40
percent \citep{Simon10}, and also affect the shape of the radial velocity
anomaly (for both the linear and the sine cases studied in this paper). This
effect could be corrected with the aid of the observational transit light
curves in reality. A statistical method to map the reconstruction of
relevant parameters of a binary planet system is beyond the scope of this
paper, but would need to be explored in details in future.

After extending the results to a hierarchical triple star system containing a
dark binary and a tertiary star, the deviation in the R-M effect would be large
enough to be detected especially if the dark binary is composed of a compact
object and a brown dwarf/planet, which may put further constraints on the
geometrical configuration of triple star systems and provide insights on their
formation and evolution.
 
We thank the referee for many helpful comments. This research was supported in
part by the National Natural Science Foundation of China under No.\ 10973001.


\begin{thebibliography}{99}

\bibitem[Albrecht et al.(2007)]{A07} Albrecht, S., Reffert, S., Snellen, I.,
Quirrenbach, A., \& Mitchell, D.\ S.\ 2007, A\&A, 474, 565

\bibitem[Carter et al.(2011)]{Carter11} Carter, J.\ A.\ et al.\ 2011, Science, 331, 562

\bibitem[Collier Cameron et al.(2010)]{CC10} Collier Cameron, A.\ et al.\ 2010,
MNRAS, 407, 507

\bibitem[Dumusque et al.(2011)]{Dumusque11} Dumusque, X., Udry, S., Lovis, C.,
Santos, N.\ C., \& Monteiro, M.\ J.\ P.\ F.\ G.\ 2010, A\&A, 525, A140

\bibitem[Fabrycky \& Tremaine(2007)]{FT07} Fabrycky, D., \& Tremaine, S.\ 2007, ApJ, 669, 1298

\bibitem[Ford et al.(2000)]{Ford00} Ford, E.\ B., Kozinsky, B., \& Rasio, F.\
A.\ 2000, ApJ, 535, 385

\bibitem[Forveille et al.(2011)]{F11} Forveille, T.\ et al.\ 2011, arXiv:1109.2505

\bibitem[Hirano et al.(2011)]{H11} Hirano, T., Suto, Y., Winn, J.\ N., Taruya,
A., Narita, N., Albrecht, S., \& Sato, B.\ 2011, ApJ, 742, 69

\bibitem[Jewitt \& Haghighipour(2007)]{JH07} Jewitt, D.,\& Haghighipour, N.\ 2007, ARA\&A, 45, 261

\bibitem[Kipping(2009a)]{K09a} Kipping, D.\ M.\ 2009a, MNRAS, 392, 181

\bibitem[Kipping(2009b)]{K09b} Kipping, D.\ M.\ 2009b, MNRAS, 396, 1797

\bibitem[Kipping et al.(2009)]{Kipping09} Kipping, D.\ M., Fossey, S.\ J., Campanella, G., Schneider, J., Tinetti, G.\ 2009, in du Foresto V.\ C., Gelino D.\ M., Ribas I.\, eds, ASP Conf.\ Ser.\ Vol.\ 430, Pathways Towards Habitable Planets. Astron.\ Soc.\ Pac., San Francisco, p.\ 139

\bibitem[Kipping et al.(2012)]{Kipping12} Kipping, D.\ M., Bakos, G.\ \'{A}.,
Buchhave L., Nesvorn\'{y}, D., Schmitt, A.\ 2012, arXiv:1201.0752

\bibitem[Kozai(1962)]{K62} Kozai, Y.\ 1962, ApJ, 67, 591

\bibitem[Li et al.(2008)]{Li08} Li, C.-H.\ et al.\ 2008, Nature, 452, 610

\bibitem[Lissauer et al.(2011)]{Lissauer11} Lissauer J.\ J.\ 2011, Nature, 470, 53

\bibitem[Lovis et al.(2010)]{Lovis10} Lovis, C., S\'{e}gransan, D., \& Mayor, M.\ et al. 2011, A\&A, 528, A112 

\bibitem[McLaughlin(1924)]{M24} McLaughlin, D.\ B.\ 1924, ApJ, 60, 22

\bibitem[Murray \& Dermott(1999)]{MD99} Murray, C.\ D., \& Dermott, S.\ F.\
1999, Solar System Dynamics (Cambridge: Cambridge Univ.\ Press) 

\bibitem[Murray-Clay \& Schlichting(2011)]{MCS11} Murray-Clay, R.\ A., \& Schlichting, H.\ E.\ 2011, ApJ, 730, 132

\bibitem[Nagasawa et al.(2008)]{NIB08} Nagasawa, M., Ida, S., \& Bessho, T.\ 2008, ApJ, 498, 508 

\bibitem[Ohta, Taruya, \& Suto(2005)]{OTS05} Ohta, Y., Taruya, A., \& Suto, Y.\ 2005, ApJ, 622, 1118

\bibitem[Podsiadlowski et al.(2010)]{P10} Podsiadlowski, P., Rappaport, S.,
Fregeau, J.\ M.\, \& Mardling R.\ A.\ 2010, arXiv:1007.1418

\bibitem[Vogt et al.(2011)]{V10} Vogt, S.\ S., Butler, R.\ P., Rivera, E.\ J.,
Haghighipour, N., Henry, G.\ W., Williamson, M.\ H.\ 2010, ApJ, 723, 954

\bibitem[Winn(2010)]{W10} Winn, J.\ 2010, Exoplanets, ed.\ S.\ Seager (Tucson,
AZ: Univ.\ of Arizona Press), (arXiv:1001.2010)

\bibitem[Winn et al.(2005)]{Winn05} Winn, J.\ N.\ et al.\ 2005, ApJ, 631, 1215

\bibitem[Winn et al.(2009)]{Winn09} Winn, J.\ N.\ et al.\ 2009, ApJ, 703, 99

\bibitem[Rossiter(1924)]{R24} Rossiter, R.\ A.\ 1924, ApJ, 60, 15

\bibitem[Sartoretti \& Schneider(1999)]{SS99} Sartoretti, P.\ \& Schneider, J.\ 1999, A\&AS, 134, 553

\bibitem[Sato \& Asada(2009)]{SA09} Sato, M.\ \& Asada, H.\ 2009, PASJ, 61, L29

\bibitem[Sato \& Asada(2010)]{SA10} Sato, M.\ \& Asada, H.\ 2010, PASJ, 62, 1203

\bibitem[Sanchis-Ojeda \& Winn(2011)]{SW11} Sanchis-Ojeda, R.\ \& Winn, J.\ N.\ 2011, ApJ, 743, 61

\bibitem[Simon et al.(2009)]{Simon09} Simon, A.\ E., Szab\'{o}, Gy.\ M., \& Szatm\'{a}ry, K.\ 2009, EM\&P, 105, 385

\bibitem[Simon et al.(2010)]{Simon10} Simon, A.\ E., Szab\'{o}, Gy.\ M., Szatm\'{a}ry, K., \& Kiss, L.\ L.\ 2010, MNRAS, 406, 2038

\bibitem[Simon et al.(2012)]{Simon12} Simon, A.\ E., Szab\'{o}, Gy.\ M., Kiss, L.\ L., \& Szatm\'{a}ry, K.\ 2012, MNRAS, 419, 164

\bibitem[Yu \& Tremaine(2001)]{YT01} Yu, Q.\ \& Tremaine, S.\ 2001, AJ, 121, 1736
\end{thebibliography}
\end{document}